\documentclass[final,5p,times,twocolumn]{elsarticle}

\usepackage[colorlinks,urlcolor=blue,linkcolor=blue,citecolor=blue]{hyperref}

\usepackage{color,array}

\usepackage{graphicx}

\usepackage{amsmath}
\usepackage{amssymb}

\newtheorem{proposition}{Proposition}
\newtheorem{remark}{Remark}

\allowdisplaybreaks

\newcommand{\tr}{\mathrm{tr}\,}

\begin{document}

\begin{frontmatter}


\title{\LARGE The Bayesian Separation Principle for Data-driven Control} 

\author{G.~Baggio$^*$, R.~Carli$^*$, R.~A.~Grimaldi$^{\circ}$, G.~Pillonetto$^*$}
\address{$^*$ Department of Information Engineering, University of Padova, Italy, \\ $^{\circ}$ Department of Electrical Engineering, Imperial College, London}
\begin{abstract}
In this paper we investigate the existence of a separation principle between model identification and control design in the context of model predictive control. First, we clarify that such a separation principle holds asymptotically in the number of data in a Fisherian context, and show that it holds universally, i.e.~regardless of the data size, in a Bayesian context. Then, by formulating model predictive control within a Gaussian regression framework, we describe how the Bayesian separation principle can be used to derive computable, uncertainty-aware expressions for the control cost and optimal input sequence, thereby bridging direct and indirect data-driven approaches. Numerical results in both linear and nonlinear scenarios illustrate that the proposed approach outperform nominal methods that neglect uncertainty, highlighting the advantages of incorporating uncertainty in~the~control~design~process.
\end{abstract}


\end{frontmatter}


\section{Introduction}

Data-driven control has emerged as a powerful paradigm in modern control systems, where the design and implementation of controllers rely directly on the information extracted from data. In contrast to classical control approaches that depend heavily on detailed mathematical models of the system dynamics, direct data-driven methods seek to bypass explicit model identification, offering greater flexibility and adaptability. These methods are particularly appealing in environments where obtaining accurate models is challenging due to complexity, uncertainty, or variability in the system. 

In recent years, several model-based control design problems have been translated into a direct data-driven control framework, including optimal linear-quadratic control \cite{DePersisTesi19, Celi23} and Model Predictive Control (MPC) \cite{coulson2019data,berberich2020data}. However, only few works have discussed the potential advantages of direct approaches over classical indirect methods, see e.g.~\cite{Dorfler2022bridging,Krishnan2021direct,Mattsson2024}. Notably, \cite{Dorfler2023b} suggests that the advantage of direct methods may stem from the lack of a separation principle between model identification and control design.

The primary objective of this paper is to shed light on the existence/non-existence of such a separation principle in the context of MPC. Of course, an answer to this question depends on the chosen probabilistic modeling framework, which defines how uncertainty in the system parameters is represented and handled. In the first part of the paper, we investigate the classic Fisherian setting \cite{Ljung:99}, where system dynamics are governed by deterministic yet unknown parameters and, for known model structure, uncertainty stems from measurement noise. We then consider the Bayesian setting \cite{SpringerRegBook2022}, where these parameters are treated as random variables and uncertainty is modeled~using~prior~and~posterior~distributions.

After establishing a precise definition of separation principle, we highlight an important yet often underappreciated result in the Fisherian context. Namely, assuming e.g.~Gaussian noises and quadratic loss functions, it is known that classic prediction error methods yield the optimal parameter estimate asymptotically in the number of data, provided that the class of parametrized models contains the true system. Such estimate in turn yields the optimal control, by virtue of the maximum-likelihood invariance principle \cite{Zhena1966}. This means that a separation principle between identification and control holds asymptotically in the number of data in the Fisherian setting. For finite data, instead, the separation principle may not hold. In contrast, the Bayesian framework inherently incorporates uncertainty about the system parameters through the posterior distribution, which encodes all available information about the model. We show that this posterior can then be leveraged to directly account for uncertainty in the MPC cost. The optimal control is then directly computed by minimizing this modified cost. Consequently, the separation principle holds universally in the Bayesian case, regardless of~the~amount~of~data.

In the second part of the paper, we delve deeper into Bayesian model predictive control, framing the problem within a Gaussian regression framework \cite{Rasmussen}. We connect it to approaches called model-free in recent control literature. In fact, even when nonlinear systems must be treated, it allows the identification step to be carried out starting from large hypothesis spaces, circumventing ill-posedness simply by including information on the input-output map regularity. This framework permits also to derive explicit expressions for the control cost in some relevant scenarios, treating the nonlinear and linear cases separately. For the nonlinear case, we show that closed-form expressions of the MPC cost can be easily derived for NFIR systems and provide a method for obtaining approximate, computationally tractable expressions for NARX systems. Instead, in the linear case, we leverage the known structure of $k$-step ahead predictors to obtain computable expressions for both the control cost and optimal input sequence. We test our proposed approach, termed Bayesian Separation Principle (BSP), in both a simple, analytically tractable linear scenario and a nonlinear case study. In both scenarios, BSP outperforms the nominal approach, which defines the MPC cost using mean parameter values and~neglect~prediction~uncertainty.

Additionally, throughout the paper, we not only clarify the definition of separation principle between identification and control, but also provide precise definitions of other frequently used yet often ambiguous terminology, such as the notions of control bias and of the aforementioned model-free control. By offering clear definitions and contextual explanations, we aim to resolve any ambiguity and foster a more accurate understanding of these concepts and their relevance in data-driven control.

To conclude, we point out that there are existing works that share similarities with ours. In particular, our work can be seen as a continuation of \cite{Scampicchio}, where Bayesian kernel-based methods have been exploited to design optimal closed-loop controllers for linear systems (for a more comprehensive overview of this and related topics, see \cite{Care}). More recently, \cite{AC-MF-VB-SF:25} framed MPC design within a Bayesian framework by focusing on the linear setting and deriving closed-form expressions for optimal inputs that account for prediction uncertainty. While similar in spirit, our work deals with the more general nonlinear setting and, even in the linear case, the control expressions we derive differ from those in \cite{AC-MF-VB-SF:25}, as detailed in the numerical section of the paper. Finally, \cite{Hewing19,Hewing20} have combined Gaussian processes with MPC to enhance control safety. Differently from our approach, in these papers the authors consider an hybrid modeling framework which combines a nominal system description with an additive Gaussian process term accommodating uncertainty.

\section{System identification and the classical approach}
The aim of system identification is to obtain a mathematical model of a dynamic system starting from input-output measurements.
An important aspect is that model construction has often to be seen as a problem of relevance. The primary interest might be to predict future data over certain operating conditions. A model might predict well one-step-ahead, e.g.~simply setting the prediction to the last measured output, but very poorly at large horizons whose limit case (infinite horizon) requires the system to be simulated without using any past output \cite{SchoukensLjung:2019}. 
Alternatively, one could ask for a model that provides a good description of the system just over frequencies of interest for the control. For MPC this could be again translated into prediction capability over a specific range of horizons, possibly using only a restricted class of inputs.

In the classical approach to system identification described in popular books like \cite{Ljung:99,Soderstrom}, the simplest scenario involves a single model  structure $\mathcal{M}$.
Examples are FIR and ARX in the linear setting or their nonlinear extensions like the NARX model which links inputs $u(t)$ (always assumed deterministic in what follows) to outputs $y(t)$ 
as follows
      \begin{equation}\label{DetMod}
        y(t) \!=\! f(y(t-1), \ldots,y(t-m), u(t),\ldots,u(t-m);\theta) + e(t)
        \end{equation}
        where $e(t)$ is white noise while the integer $m$ defines the system memory.
The model depends on an unknown (finite-dimensional) parameter vector $\theta$. The resulting parametrized structure $\mathcal{M}(\theta)$ has to be estimated from  
$N$ input-output data $\{u(t),y(t)\}_{t=1}^N$ collected during an experiment and contained in a set denoted by $\mathcal{D}$. Assume for the time being  $y(t)$ scalar with $\hat{y}(t|t-1;\theta)$ to indicate its one-step-ahead predictor induced by the model. This predictor depends on all the outputs up to instant $t-1$ and corresponds to $f(y(t-1), \ldots, u(t), \ldots;\theta)$ when the NARX model \eqref{DetMod} is adopted.  The unknown parameter vector can then be estimated
via the prediction error minimization (PEM) approach: 
\begin{equation}\label{PEM}
\hat{\theta} = \arg\min_{\theta} \sum_{t=1}^N
\big(y(t)-\hat{y}(t|t-1;\theta)\big)^2. 
\end{equation}
The performance of this estimator 
can be measured in terms of the mean squared error (MSE). The MSE is computed before seeing
the identification data, with $\hat{\theta}$ therefore being interpreted as a random vector because of the dependence on the stochastic outputs $y(t)$. 
Using $\mathbb{E}$ to indicate the expectation operator, it is given by the following function of $\theta$: 
\begin{eqnarray}\label{MSEch1}
\text{MSE}_{\hat{\theta}}(\theta) &=& \mathbb{E}\| \hat{\theta} - \theta \|^2 \nonumber \\
&=&  \underbrace{\sum_{i}  \mathbb{E}  (\hat{\theta}_i - \mathbb{E} \hat{\theta}_i )^2}_{\text{Variance}} 
+   \underbrace{\sum_{i}  (\theta_i - \mathbb{E} \hat{\theta}_i )^2}_{\text{Bias}^2},
\end{eqnarray}
where $\theta_i$ is the $i$th component of $\theta$. The last equality points out the error decomposition 
into its variance and bias component. If
$$
\mathbb{E}[\hat{\theta}]=\theta \quad \forall \theta
$$
the estimator has null bias component and is therefore called unbiased.\\

PEM has a long and fruitful history in system identification. Under Gaussian noises and identifiability assumptions,  the PEM estimator enjoys some optimality properties. Its link with maximum likelihood (ML) in fact guarantees that it cannot be outperformed by any other unbiased estimator as the data set size grows to infinity: the Cramer-Rao bound will be reached making PEM variance as small as possible  \cite{CaseBerg:01}. Such property also 
transfers to the estimate of any function of $\theta$ just replacing the unknown parameter vector with its PEM estimate by virtue of the 
ML invariance principle \cite{Zhena1966}. We will come back to talk extensively around this point later in the MPC context.\\

The classical approach and its properties are Fisherian in nature: one postulates that there exists a true deterministic system $\mathcal{S}$ that has generated the data and tries to learn it using only the experimental data. 
Asymptotic optimality of the estimation procedure then holds assuming that the postulated structure $\mathcal{M}$ contains the true system $\mathcal{S}$.
However, this scenario is often too simplified. Also model complexity
is typically unknown, in particular the dimension of $\theta$ which is related to the system memory $m$ in \eqref{DetMod}.  More than one structure needs to be introduced and criteria such as AIC or BIC are used to determine the most appropriate one. Such criteria try to find a good trade-off between the two MSE components reported in \eqref{MSEch1}. In fact,\\ 

\noindent \emph{more parameters provide more model flexibility and allow a better fit of the identification data, decreasing the bias. But more parameters also allow adjustments to the noise in the
estimation data, augmenting the variance term. The improved fit due to such adjustments has no value and can decrease model prediction capability\footnote{While this classical principle of parsimony suggests that the model should describe the experimental data without being too complex,
there are however some overparametrized architectures, like deep networks, that only emphasize fitting data and may still perform well on future (unseen) data. For a recent survey on such phenomenon the reader may refer to \cite{DeepSurvey2025}.}.}\\
 
Assuming known (for the sake of simplicity) the variance $\sigma^2$ of the stochastic noise $e(t)$ which influences the system, 
the final model returned by the identification procedure is $\mathcal{M}(\hat{\theta})$ where now 
\begin{align}\label{FinalM} 
\hat{\theta}  = \arg\min_{\theta} \sum_{t=1}^N
\frac{\big(y(t)-\hat{y}(t|t-1;\theta)\big)^2}{\sigma^2}\notag + J(N)\dim(\theta)\\ 
\end{align}
with optimization performed also w.r.t.~the dimension of $\theta$. 
Interestingly, the choice of the penalty term $J$ in \eqref{FinalM} is also a problem of relevance at least in terms of asymptotic properties \cite{Zhang2023}. AIC, obtained using $J(N)=2$, enjoys minimax properties described in \cite{Yang2005} that may lead to advantages in prediction. However, it overestimates the model order with finite probability. BIC, defined by $J(N)=\log(N)$, is not minimax but asymptotically will return the correct order assuming that this latter exists (parametric scenario). 
For an in-depth discussion on the properties of estimators related to model order selection see also \cite{Leeb2005}.\\

\section{MPC: direct and indirect approach}

The arguments developed in this paper concern two distinct phases of an experiment involving a dynamic system: identification, described in the previous section, and control, introduced below. For what regards notation, in what follows it is convenient to use $\tau$ as the starting time of the control phase and assume that the identification data set $\mathcal{D}$ contains input-output pairs observed before $\tau$. For a generic time instant $t$, we also use $y^{t-1}$ to denote the system outputs up to $t-1$. When used as an argument of the NARX system $f$ in \eqref{DetMod} this has however to be understood as the vector containing only the previous $m$ outputs. Similar considerations apply to the notation $u^t$.\\

Consider now MPC where, in the first instance, the system $\mathcal{S}$ is assumed perfectly known. In addition, the deterministic input $u$ is manipulable.
After observing $y^{\tau-1}$, the problem is to calculate future inputs $u(\tau),u(\tau+1),\ldots$ that allow the future outputs $y(\tau),y(\tau+1),\ldots$ to follow a reference $y_{\text{ref}}(\tau),y_{\text{ref}}(\tau+1),\ldots$. Knowledge of $\mathcal{S}$ allows us to define the optimal $\ell$-step ahead predictors $\hat{y}(\tau+\ell|t)$  by setting them to the mean of future outputs. 
In this way the predictors minimize the MSE on output prediction. 
Using quadratic losses, the ideal control cost function to be minimized w.r.t.~$u$ over an horizon of length $T$ is formulated as: 
 \begin{align}\label{ControlCost}
  J_u  \! &= \!  \sum_{\ell=0}^{T-1}\|y_{\text{ref}}(\tau+\ell) - \hat{y}(\tau+\ell)\|_{Q_\ell}^2\! +\! \sum_{\ell=0}^{T-1}\|u_{\text{ref}}(\tau+\ell) - u(\tau+\ell))\|_{R_\ell}^2
        \end{align}
        where here, and in what follows, $\hat{y}(\tau+\ell)$ is used in place of
      $\hat{y}(\tau+\ell|\tau)$ to simplify notation.  
In \eqref{ControlCost} the outputs can now be also vectors, $Q_\ell$ and $R_\ell$ are positive semidefinite matrices, and $\|x\|_M^2:=x^\top M x$ for any matrix $M$. Finally, $u_{\text{ref}}$ are reference inputs that define the penalty on the energy required by the control.\\ 

In a more realistic MPC scenario $\mathcal{S}$ is not known and system identification can enter the scene first. Following our previous discussion, we can introduce a parametric model structure $\mathcal{M}(\theta)$ to capture $\mathcal{S}$. All the
induced predictors become $\hat{y}(\tau+\ell;\theta)$, function of the unknown vector $\theta$. In turn, this
leads to the parametrized control cost $J(u;\theta)$. 
In this setting the simplest version of the so-called \emph{indirect approach} to data-driven control, of which the classical identification mentioned above is just an example, first determines $\theta$ from the input-output data $\mathcal{D}$ collected in the previous experiment. This can be done through PEM, i.e.~minimizing the one-step-ahead prediction error as in \eqref{PEM}, possibly solving the more complex problem \eqref{FinalM} in the case of unknown model complexity.
Then, the objective $J(u;\theta)$ is minimized by replacing $\theta$ with its estimate (the realization of $\hat{\theta}$).\\

There are some recurring criticisms of the indirect approach that have emerged in recent years. The first one is that bias is introduced in the control objective \cite{Dorfler2023}. 
Note however that such issue depends on the indirect approach adopted. For instance, classical system identification is parametric in nature but one could also consider different non parametric and/or Bayesian settings as we will do later on. Furthermore a careful definition of bias is needed. 
To this regard,\\

\noindent\emph{in the Fisherian context where $\theta$ is deterministic, we say that the indirect approach to control is unbiased if, for any input $u$, one has
$$
\mathbb{E}[J(u;\hat{\theta})]=J(u;\theta) \quad \forall \theta,
$$
with expectation taken w.r.t.~the distribution of the estimator~$\hat{\theta}$.}\\

\noindent Assume that the model structure $\mathcal{M}$ contains $\mathcal{S}$ (so that there exists $\theta$ leading to the optimal predictors). Then, even if $\theta$ is unknown, null bias guarantees that the mean of the cost defined using $\hat{\theta}$ corresponds to the ideal objective \eqref{ControlCost}.\\

Adopting the classical PEM approach, the unbiasedness property hardly holds because $J$ is a nonlinear function of $\theta$.  
On the other hand, when e.g.~the identification data $\mathcal{D}$ are nonlinearly related to the model parameters, $\hat{\theta}$ is rarely an unbiased estimator of $\theta$ itself\footnote{This is by no means in itself a negative thing. We are here concerned with unbiased costs but there exist estimators that introduce a small bias to greatly decrease variance, hence leading to a favourable MSE. Stein also proved in the linear regression setting that, if the dimension of $\theta$ exceeds two, the least squares estimator is not admissible: there are biased estimators that have smaller MSE for any $\theta$, e.g. see \cite{Efron1973} and \cite[Chapter 1]{SpringerRegBook2022}.}.
The rare exceptions include linear regression problems with linear least squares adopted  to estimate $\theta$. 
For this reason one could try to remove this distortion exploiting the experimental observations in a different way. Indeed, the idea that permeates the \emph{direct approach} is that the information contained in the data should be maximally exploited by focusing directly on the control cost. So,\\

\noindent\emph{the problem of relevance related to the direct approach to data-driven control is to build directly from input-output data an estimator of (ideal) control objectives like \eqref{ControlCost} having desired statistical properties (like unbiasedness).}\\

Another argument against the indirect approach is that there does not exist a separation principle linking system identification to control. It is said that the “optimal” solution to the control cannot be achieved by first solving the identification problem and then, in cascade, using the result for control purposes.
But, as in the case of bias, also this statement may not be correct if the system identification setting is not precisely defined. Indeed, we will see that a separation principle actually exists if we move within a more modern paradigm of identification which connects dynamic systems, machine learning, and Bayesian regularization. 

\section{Separation principle for identification and control}

Linking to the discussion given at the end of the previous section, first it is important to provide a definition of separation principle. We say that\\ 

\noindent\emph{a separation principle for identification and control holds if the optimal solution coming from the system identification step allows also to obtain the optimal solution to the control problem.}\\

Again, it is important to emphasize that the existence 
of this principle depends on the context in which we work, the stated assumptions, and also the definition of optimality. In particular\\

\noindent\emph{for the control problem, optimality will henceforth be understood as the definition of the best possible estimator of the MPC cost according to some useful metric, e.g.~the mean squared error. Hence, we say that the control input $u$ is  optimal if it minimizes the optimal estimate of the MPC cost.}

\subsection{Fisherian setting}

The mean squared error introduced in \eqref{MSEch1} measures the
performance of the estimator $\hat{\theta}$, function of the identification data $\mathcal{D}$, in reconstructing $\theta$.
Such notion immediately extends to measuring the distance between any function $g(\theta)$ and $g(\hat{\theta})$ as follows
$$
\text{MSE}_g(\theta) = \mathbb{E}\big[(g(\theta)-g(\hat{\theta}))^2\big].
$$
Remembering its dependence on the unknown $\theta$, 
it is well known that it is not possible to obtain an optimal estimator in terms of MSE considering all possible functions of the data. The simplest example is the constant estimator which always returns the same estimate $\bar{\theta}$ independently of the measurements in $\mathcal{D}$.
It will outperform any other estimator if the true parameter vector is equal to $\bar{\theta}$ but its MSE can rapidly increase moving to other regions of the parameter space. For MPC, where the objective to estimate is $J(u;\theta)$, this means that it is not even possible to define the optimal solution adopting the direct approach to the control. A fortiori, we conclude that in the Fisherian setting a separation principle cannot exist using the MSE as optimality criterion if we work inside the class of all the possible estimators.\\

A separation principle instead arises if we restrict the solutions to the class of unbiased estimators and reason on an asymptotic basis. It is derived from the theory of maximum likelihood (ML) estimators. In particular, we have already stressed that PEM concurs with the ML estimator of $\theta$, here denoted by $\hat{\theta}^{ML}$, assuming e.g. Gaussian noises and quadratic losses as done in \eqref{PEM}. Given any $g(\theta)$, the invariance principle then states that its ML estimator is obtained just replacing $\theta$ with $\hat{\theta}^{ML}$. This already hints at why the MPC solution obtained by exploiting this (preliminary) system identification step is asymptotically optimal. In fact, we can carry out such replacement in the ideal Fisherian objective \eqref{ControlCost} obtaining $J(u;\hat{\theta}^{ML})$ which thus represents the maximum likelihood estimator of the control cost. Hence, if  the postulated structure contains the true system $\mathcal{S}$, 
as the data size goes to infinity $J(u;\hat{\theta}^{ML})$ becomes optimal inside the class of unbiased estimators. Its mean converges to $J(u;\theta)$ for any $\theta$ and no other unbiased estimator of \eqref{ControlCost} can asymptotically have smaller variance. This argument still holds if more than one structure is postulated and BIC is used to control model complexity in \eqref{FinalM}  since we have seen that the exact dimension of $\theta$ is asymptotically returned. Of course, the limitation of this result relies in its asymptotic nature. 

An important point for real applications (that will come naturally in the Bayesian context) is to take into account also the uncertainty around the nominal model $\hat{\theta}^{ML}$ and then around the MPC cost. This aspect of the problem also gave rise to the rich literature on robust identification. It aims to derive confidence bounds useful e.g.~to design robust controllers, capable of stabilizing a whole set of systems that contains $\mathcal{S}$ with high confidence, see e.g. \cite{Goodwin1992,Ljung:2014,Baggio2022}. 

\begin{remark}
The separation principle here illustrated should not be misunderstood with the concept of sufficient statistic also discussed in \cite{AC-MF-VB-SF:25}. Given a model parametrized by $\theta$, a sufficient statistic
$T(\mathcal{D})$ 
is a transformation of the data set $\mathcal{D}$ able to retain all the information contained in $\mathcal{D}$ for estimating $\theta$. This manifests itself in the fact that the MSE performance of any estimator of $g(\theta)$, which is modified by conditioning it on $T(\mathcal{D})$, improves or remains the same for any $\theta$. It follows that, if the minimum variance unbiased estimator of the MPC cost exists, it can be written as function only of $T(\mathcal{D})$.
So, the construction of a sufficient statistic provides optimal 
compression 
of the data to estimate \eqref{ControlCost}.  However, the mere construction of $T(\mathcal{D})$  does not define or even guarantee the existence of the optimal estimator. Hence, it does not lead to fulfilling a separation principle according to the definition given above.   
\end{remark}

\subsection{Bayesian parametric setting}

We will now see that the separation principle applies in the Bayesian setting regardless of the data set size. This is first discussed in a parametric scenario where $\theta$ is still finite dimensional but now represents a random vector.

Our assumption (which will be relaxed later) is that the system can still be described by a parametric structure $f$. But now it depends on the random (instead of deterministic) vector $\theta$ of known probability density function (pdf) $p(\theta)$. Output data are generated as follows
      \begin{equation}\label{StochMod}
        y(t) = f(y^{t-1}, u^t;\theta) + e(t)\\ 
        \end{equation}
where 
$e(t)$ is white noise, independent of $\theta$, whose pdf is known. In this Bayesian setting, the solution of the identification problem 
has to be seen as the entire posterior of $\theta$, calculated by the Bayes rule as follows
$$
p(\theta|\mathcal{D}) = \frac{p(\mathcal{D}|\theta)p(\theta)}{p(\mathcal{D})}.
$$
In the following we use $\theta^P$ to denote $\theta$ conditional on $\mathcal{D}$, whose pdf is therefore given by $p(\theta|\mathcal{D})$.\\

Let us consider MPC in this Bayesian setting. 
The first important issue is to obtain the counterpart of the Fisherian objective \eqref{ControlCost}.
The ideal control cost function to be minimized w.r.t. $u(\tau),u(\tau+1),\ldots$ is now a stochastic variable, given by
 \begin{align}\label{ControlCostBayes}
  J_u(\theta,E)   &=   \sum_{\ell=0}^{T-1}\|y_{\text{ref}}(\tau+\ell) 
  - f(y^{\tau+\ell-1},u^{\tau+\ell};\theta)\|_{Q_\ell}^2 \nonumber\\ 
  &+ \sum_{\ell=0}^{T-1}\|u_{\text{ref}}(\tau+\ell) - u(\tau+\ell))\|_{R_\ell}^2,
        \end{align}
whose randomness comes from the dependence on $\theta$ and on the future noises $e(\tau),e(\tau+1),\ldots$ contained in $E$. 
In fact, the outputs in $y^{\tau-1}$ have not to be seen as stochastic: they have been already observed and set to their realizations. 
The future outputs $y(\tau),y(\tau+1),\ldots$ are instead stochastic and are
generated by $\theta$ and the future noises through \eqref{StochMod}.
Together with $y^{\tau-1}$, they define recursively all the optimal predictors through
$$
\hat{y}(\tau+\ell) = f(y^{\tau+\ell-1},u^{\tau+\ell};\theta).
$$
Note that, differently from the Fisherian case, the predictors depend explicitly on the future outputs (and are therefore random variables). They are not defined by computing their means because this would remove the dependence on $\theta$ leading us to a control objective far from the ideal one.

The bond between Bayesian identification and data-driven control comes about in a very natural way. Since $\theta$ and the future noises are not accessible to measurement,  to obtain an estimator  of \eqref{ControlCostBayes} first we  condition $J_u$ 
on the identification data $\mathcal{D}$, obtaining the following key random variable 
$$J_u^P:=J_u\,|\,\mathcal{D}$$
which corresponds to the conditional cost.
Then we take its expectation obtaining the estimator $\hat{J}_u$ function only of $\mathcal{D}$ (and the control input).

Since the future noises in $E$ are independent of 
$\mathcal{D}$, the conditional cost $J_u^P$ actually depends only on $\theta^P$ and $E$. Let us define the conditional predictor at instant $\tau$ as follows
$$
\hat{y}^P(\tau) := f(y^{\tau-1}, u^\tau;\theta^P) 
$$
so that $y(\tau)$ conditional on $\mathcal{D}$ is
$$
y^P(\tau):=\hat{y}^P(\tau)+e(\tau). 
$$
The next conditional predictor is
$$
\hat{y}^P(\tau+1) := f(y^P(\tau),y^{\tau-1}, u^{\tau+1};\theta^P) 
$$
so that the next conditional output is
$$
y^P(\tau+1):=\hat{y}^P(\tau+1) + e(\tau+1) 
$$
and so on. 
We can now write the conditional MPC cost as
\begin{align}\label{ControlCostCond}
  J^P_u  &=  \sum_{\ell=0}^{T-1}\|y_{\text{ref}}(\tau+\ell) - \hat{y}^P(\tau+\ell)\|_{Q_\ell}^2 \nonumber\\ 
  &+ \sum_{\ell=0}^{T-1}\|u_{\text{ref}}(\tau+\ell) - u(\tau+\ell))\|_{R_\ell}^2
\end{align}
whose posterior mean is 
\begin{equation}\label{PosteriorAvgCost}
\hat{J_u}= \int J^P_u(\theta,E) p(\theta| \mathcal{D}) p(E) d\theta dE. 
\end{equation}      
Bayesian estimation theory \cite{Anderson:1979} ensures that this function $\hat{J_u}$ of the identification data $\mathcal{D}$ minimizes the mean squared error from $J_u(\theta,E)$ 
over the class of all possible estimators. Furthermore, $\hat{J_u}$ is unbiased and gives the minimum variance estimate of the MPC cost. This means that in the Bayesian scenario the question if indirect is better or not than direct 
has found a simple answer: nothing can be better than computing the posterior, using it to average the uncertainty around the MPC objective and obtaining the control input as the minimizer of the resulting estimate $\hat{J_u}$. So,\\

\noindent\emph{in the Bayesian setting the posterior of $\theta$ is the optimal solution of the identification step and contains all the information needed for optimal estimation of the MPC cost. Hence, the separation principle holds.}\\

\begin{remark}\label{RemSepPr}
It is easy to see that if we used other loss functions in the MPC cost, such as $\ell_1$ penalties instead of quadratic ones, the statement above would still hold. In particular, the separation principle continues to hold for any other MPC cost that depends only on $\theta$ and future noises $E$. 
\end{remark}

\begin{remark}\label{RemIrrNoise}
    The cost \eqref{ControlCostBayes} could be also reformulated replacing each predictor
    $f(y^{\tau+\ell-1},u^{\tau+\ell};\theta)$ with the output $y(\tau+\ell)$.
    But nothing substantial would change: the posterior mean $\hat{J_u}$ would just contain some additional irreducible error terms given by variances of the future noises in $E$. 
\end{remark}

\subsection{Bayesian nonparametric setting: Gaussian regression}
\label{KerModels}

The solution to the MPC problem derived in the previous setting could not yet be satisfactory under some aspects. We have seen that the dimension of $\theta$ is often unknown in real applications and its estimation is not trivial. Criteria such as AIC or BIC, very much in tune with this Bayesian setting, have also shown some weaknesses to control complexity \cite{PillonettoDCNL:14}. The formulation of a  finite-dimensional structure with good prediction capability can therefore be  difficult as well as the definition of a suitable prior on $\theta$.
This kind of criticism is very much in line with the desire of direct approaches to break free from the model in some way. In the MPC literature many of the proposed techniques are called (or aspire to be) \emph{model-free}. This is somewhat utopian since in practice a data model needs always to be introduced, at least implicitly. In  statistical terms, the term model-free should be connected with the use of nonparametric structures leading to null (or negligible) model bias. The latter has not to be confused with the estimation bias mentioned earlier. Specifically, we say that\\  

\noindent\emph{the approach is model-free or, equivalently, nonparametric if it has negligible model bias. This means that the system is searched in a space of such large size (possibly even infinite dimensional) that it contains virtually all possible systems and predictors of interest.}\\

One way to get a Bayesian model-free approach to MPC is to cast Gaussian regression \cite{Rasmussen} into the control framework. 
There are many different ways of introducing this estimation technique. For example, Neal in \cite{Neal1995} considered a neural network with independent and random weights and showed that the input-output map becomes a Gaussian process as the width of the network grows to infinity.
This construction illustrates well the richness that this type of model can have. Similarly, below we introduce Gaussian regression starting from a wide range of basis functions whose coefficients are random variables.\\

First, our stochastic system $f$ in \eqref{StochMod} is seen as the sum of a large, possibly infinite, number of (deterministic) basis functions
$\phi_k$. Their stochastic  coefficients are the components of  $\theta$, modeled as zero-mean independent 
Gaussian random variables of variance $\zeta_k$, i.e.
\begin{equation}\label{Priorth}
    \theta_k \sim \mathcal{N}(0,\zeta_k).
\end{equation}
It is convenient to streamline the notation for the argument of $f$, called also input location in machine learning, letting
$$
x=[y^{t-1} \ \ u^t]^\top
$$
where, in the case of NARX models, 
$y^{t-1}$ and $u^t$ going back in time only to the instant $t-m$.
Now, we can write 
\begin{equation}\label{ExpPhi}
f(x)= \sum_{k=1}^p \theta_k \phi_k(x)
\end{equation}
where $p$ could also grow to infinity.
To further simplify notation, we also assume that $f$ is real-valued and, just for a while, that $x$ is scalar. This permits to introduce some simple and illustrative basis functions like
sinusoids
\begin{equation}\label{SplineBF}
\phi_k(x) = \sin(x(k\pi-\pi/2)),
\end{equation}
monomials
\begin{equation}\label{MonomialsBF}
\phi_k(x) = x^k,
\end{equation}
and a variation of them obtained by multiplication with radial functions such as the squared exponential:
\begin{equation}\label{GaussianBF}
\phi_k(x) = \exp\left(-\frac{x^2}{\eta}\right) x^k,
\end{equation}
where $\eta$ is called kernel width.
A common feature of these $\phi_k$ is that they are universal, i.e., they are able to approximate arbitrarily well any continuous function on any compact set. However, it is not possible to learn a very large/infinite number of coefficients from the finite number of data contained in $\mathcal{D}$. This intrinsic ill-posedness is overcome by the prior \eqref{Priorth}, whose aim is to specify some high-level information about the system. In particular, kernel-based (Gaussian) regression is obtained by specifying how quickly the variances $\zeta_k$ decay to zero, so that
\begin{itemize}
    \item the expected level of system smoothness is encoded in $f$ leading to a prior model capable of describing practically any (continuous) input-output relationship;
    \item the covariance of $f$ given by
    $$
    K(x,a):=\mathbb{E}[f(x)f(a)],
    $$
    also called kernel, becomes available in closed-form. 
\end{itemize}
Fulfillment of the second point leads to important computational advantages. No basis functions need to be stored in memory being all implicitly embedded in $K$, a point related to the so-called kernel trick \cite{Scholkopf01b}.  It allows the efficient solution of a wide variety of nonlinear optimization problems for regression and classification.\\ 

Using the sine waves \eqref{SplineBF}, the choice of the variances
in \eqref{Priorth} given by
\begin{equation}\label{Decay1}
\zeta_k \propto (k\pi-\pi/2)^{-\alpha}, \quad \alpha=2
\end{equation}
defines the first-order spline kernel
\begin{equation}\label{Spline}
K(x,a)= \min(x,a)
\end{equation}
which includes just information on continuity of $f$ \cite{Wahba:90}.
Increasing the value of $\alpha$ in \eqref{Decay1} augments 
expected smoothness leading, e.g., to the second-order spline kernel \cite{Bell2004}. These two covariances are likely the most famous to estimate functions over one-dimensional domains.

Adopting \eqref{GaussianBF} and further increasing the decay rate of the variances as follows
\begin{equation}\label{Decay2}
\zeta_k \propto \frac{2^k}{\eta^k k!}
\end{equation}
defines the widely used Gaussian kernel \cite{Scovel10,Minh:09}. It is infinitely differentiable and suited to describe functions particularly smooth. In the general multivariate case, where $x$ and $a$ can now be vectors containing past inputs and outputs, it assumes the form
\begin{equation}\label{GK}
K(x,a) = \exp\left(-\frac{\|x-a\|^2}{\eta}\right).
\end{equation}

A significant variation described in \cite{Pillonetto:11nonlin} makes use of a diagonal matrix $D$ that weights the influence of the components of the input locations differently:
\begin{equation}\label{GKD}
K(x,a) = \exp\big(-(x-a)^\top D (x-a)\big).
\end{equation}
This leads to a form of physics-informed machine learning which makes use of regularizers which incorporate information on dynamical systems.
In particular, the concept of fading memory can be included  by partitioning $D$ into two diagonal blocks $D_y$ and $D_u$ one dedicated to outputs, the other to inputs, whose components tend to zero. For example,
the entries along the diagonal of $D_y$ (or of $D_u$) can be defined by 
$$
[D_y]_{ii} \propto \alpha^i \ \ \mbox{with} \  \ 0< \alpha <1 
$$
to model exponential decay with the rate $\alpha$. 
The kernel thus takes into account the fact that the influence of past inputs and outputs on the current output disappears over time.

Another important covariance is the polynomial kernel \cite{Poggio75}
which includes all the monomials \eqref{MonomialsBF} up to order $r$:
\begin{equation}\label{PK}
K(x,a):= \big(x^\top a +1\big)^r
\end{equation}
Hence, \eqref{PK} includes implicitly a number of basis functions which scales exponentially with $r$ and the dimension of $x$.\\

Finally, if the predictor is linear in the past inputs and outputs,
we can rewrite \eqref{StochMod} in matrix-vector form obtaining
     \begin{equation}\label{StochModLin}
        Y = F \theta + E\\ 
        \end{equation}
where $Y$ is the vector collecting all the output data in $\mathcal{D}$, $F$ has two Toeplitz matrices side by side with past input and output data  and $\theta$ now indicates the (column) random vector with the predictor impulse responses coefficients. 
In this case, one can use 
a linear kernel 
\begin{equation}\label{LinKer}
K(x,a)= x^\top M a
\end{equation}
and one can easily see that the matrix $M$ now represents the covariance of $\theta$. Similarly to what discussed about the matrix $D$ entering the nonlinear kernel \eqref{GKD}, $M$ can be partitioned into two blocks $M_y$ and $M_u$, dedicated to outputs and inputs. Information on smooth exponential decay can then be introduced adopting the class of stable, TC or DC kernels reviewed in  \cite{PillonettoDCNL:14,PillonettoPNAS}. For instance, the $(i,j)$ entry of $M_y$ (or of $M_u$) using TC is 
$$
[M_y]_{i,j} = \alpha^{\max(i,j)}, \ 0 \leq \alpha <1
$$
where $\alpha$ regulates the decay rate.\\

In this section we have thus seen that in the Bayesian setting $f$ is seen as a random surface and\\

\noindent \emph{kernels are a way to define a nonparametric (model-free) approach
where the system is a zero-mean Gaussian random field of covariance $K$: 
\begin{equation}\label{GRF}
f \sim \mathcal{N}(0,\lambda K)
\end{equation}
with $\lambda$ being a positive scalar. 
}\\ 

It is now necessary to study the shape of the MPC cost resulting from such a modeling choice.

\section{MPC cost using Gaussian regression}

An issue that could weaken the Bayesian separation principle is that direct approaches also wish to provide control objectives of simple structure, possibly depending on few parameters that can be tuned from data. Given a control input $u$, the evaluation of the Bayesian control objective $\hat{J_u}$ instead requires the solution of the integral \eqref{PosteriorAvgCost} whose computational cost could be high, or even prohibitive for real-time control applications. 
Thus, it is now important to study the estimator $\hat{J_u}$ of the MPC cost reported in \eqref{PosteriorAvgCost} within the Gaussian regression framework, also understanding what parameters it depends on and how they can be estimated from the data.

\subsection{Inference using Gaussian regression}

In what follows, just to simplify exposition, the system output is assumed scalar, so that $f$ is a priori described as a real-valued Gaussian random field.  According to the Bayesian separation principle, first we need to obtain the posterior of $f$ conditional on the input-output measurements taken before the time $\tau$ when the control phase starts.

It is convenient to build from the identification data contained in 
$\mathcal{D}$ the input locations
\begin{equation}\label{zt}
z^t=[y^{t-1} \ \ u^t]^\top, \quad t=1,\ldots,N,   \end{equation}
a notation that makes them ``dynamic" in light of the dependence on time $t$\footnote{Strictly speaking, these input locations contain also some past data not contained in $\mathcal{D}$, e.g. input and outputs measured before $t=1$. If such data are not available one can handle initial conditions effects as e.g. described in \cite[Section 3.2]{Ljung:99}.}.
Recall that the $z^t$ are generated according to \eqref{StochMod} with $f$ that satisfies \eqref{GRF} and is independent of the noises $e(t)$ which form a white Gaussian noise of variance $\sigma^2$.
In the next proposition $z^*$ can be interpreted as a future input location where $f$ needs to be predicted to evaluate the MPC cost. The formulas given may seem standard, coming from basic results regarding jointly Gaussian vectors estimation \cite{Anderson:1979}. Actually,  there are some proof-related subtleties discussed in Appendix related to the fact that output data are also part of the input locations.

\begin{proposition}\label{Prop1}
Let $\mathbb{K}$ denote the $N \times N$ kernel matrix whose $(i,j)$ entry is $K(z^i,z^j)$ and let $\Sigma=\mathbb{K} + \gamma I$. Define also the weights vector $c$ and the regularization parameter $\gamma$ as follows:
$$
c= \Sigma^{-1}Y, \quad \gamma=\frac{\sigma^2}{\lambda}
$$
where the $N$-dimensional vector $Y$ contains the outputs in $\mathcal{D}$.
Then, $f$ conditional on $\mathcal{D}$ remains Gaussian. In addition, the posterior mean and variance at $z^*$ are given, respectively, by
\begin{eqnarray}\label{EandV1}
    \mathbb{E}[f(z^*)|\mathcal{D}]&=& \sum_{i=1}^N c_i K(z^*,z^i)\\
    \mathrm{Var}[f(z^*)|\mathcal{D}] &=& \lambda K(z^*,z^*) \label{EandV2}
    - \lambda \Gamma  \Sigma^{-1} \Gamma^\top
\end{eqnarray}
where
$$
\Gamma=[K(z^*,z^1) \ \ldots K(z^*,z^N)].
$$
If data are generated according to the linear model 
\eqref{StochModLin} and the linear kernel \eqref{LinKer} is
adopted, the posterior mean and variance of the one-step ahead predictor coefficients in $\theta$ are given by
\begin{eqnarray}\label{EandV1Lin}
    \mathbb{E}[\theta | \mathcal{D}]&=& \big(F^\top F+ \gamma M^{-1}\big)^{-1}F^\top Y  \\
    \mathrm{Var}[\theta|\mathcal{D}] &=& \sigma^2 \big(F^\top F+ \gamma M^{-1}\big)^{-1}.
\end{eqnarray}
Finally, given $u$, in the linear or nonlinear case the (marginal) probability density function of $Y$ is 
\begin{equation}\label{ML1}
p(Y) = \frac{\exp\left(-\frac{1}{2} Y^\top
\Sigma_y^{-1} Y \right)}{\sqrt{\det(2\pi\Sigma_y)}}, \quad \Sigma_y=\lambda \Sigma. 
\end{equation}
\end{proposition}

The above proposition thus provides the predicted output and also its posterior variance through \eqref{EandV1} and \eqref{EandV2}. However, these formulas depend on parameters that may be unknown, particularly the noise variance $\sigma^2$ and the kernel scaling factor $\lambda$
(and possibly others like the kernel width in the Gaussian kernel).
These variables, often called \emph{hyperparameters}, can be estimated via the empirical Bayes approach by maximizing the marginal likelihood given by \eqref{ML1} \cite{Hastie01}. This tuning approach enjoys important properties. In particular, while the numerator in \eqref{ML1} accounts for data fit, the denominator represents an Occam's factor which automatically penalizes too complex models \cite{MacKayNN}. To analyze the structure of the ML estimates it is useful to introduce the concept of degrees of freedom $q(\gamma)$. Using $H(\gamma)$ to denote the so-called hat matrix, defined by
\begin{equation}\label{HatM}
H(\gamma)=\mathbb{K}(\mathbb{K} + \gamma I)^{-1}, \quad \gamma=\frac{\sigma^2}{\lambda},
\end{equation}
they are given by the trace of this matrix, i.e.
\begin{equation}
    q(\gamma) = \tr\big[H(\gamma)\big].
\end{equation}
As $\gamma$ varies from $+\infty$ (strongest regularization) to 0 (absence of regularization), $q(\gamma)$ varies from 0 to the number of data $N$, hence representing a normalized measure of model complexity. After introducing the vector
$$
\hat{f} = H(\gamma)Y,
$$
the sum of squared residuals and of weighted squared estimates are, respectively, given by
$$
\text{WSRR}(\gamma)= \|Y-\hat{f}\|^2
$$
and
$$
\text{WSSU}(\gamma)= \hat{f}^\top \mathbb{K}^{-1}\hat{f}.
$$
It then turns out that the ML estimates of $\lambda$ and $\sigma^2$
satisfy \cite{DeN1997}
\begin{equation}\label{lambda}
    \lambda = \frac{\text{WSSU}(\gamma)}{q(\gamma)}
\end{equation}
and
\begin{equation}\label{sigma2}
\sigma^2 = \frac{\text{WSRR}(\gamma)}{N-q(\gamma)}.
\end{equation}
Combining these two equations, the estimate of $\gamma$ solves
\begin{equation}\label{gamma}
\gamma \frac{\text{WSSU}(\gamma)}{q(\gamma)}=\frac{\text{WSRR}(\gamma)}{N-q(\gamma)}
\end{equation}
while, for known $\sigma^2$, it solves $\gamma\text{WSSU}(\gamma)=\sigma^2 q(\gamma)$. It then follows that\\

\noindent \emph{in Gaussian regression the posterior of $f$, and hence the MPC cost induced by the Bayesian invariance principle, depends only
on the regularization parameter $\gamma$ which can be determined simply by solving \eqref{gamma}.}

\begin{remark}
The sole dependence on $\gamma$ is due to the fact that, even in the case of unknown noise variance, the latter is linked to $\gamma$ via 
\eqref{sigma2} (and then $\lambda=\sigma^2/\gamma$). Note also that \eqref{sigma2} deeply recalls noise variance estimation
in linear parametric regression \cite{CaseBerg:01} where the degrees of freedom correspond to the dimension of the unknown parameter vector. In that setting, the correction given by subtracting the number of parameters from the number of data ensures unbiasedness of the variance estimator.
\end{remark}

\subsection{Explicit expressions of MPC cost}

We derive here some explicit expressions of the MPC cost \eqref{PosteriorAvgCost}, by considering the Gaussian regression framework illustrated before. We treat separately the case of linear and nonlinear models and, to simplify notation, we let $y^{\text{ref}}_{\tau+\ell}:=y_{\text{ref}}(\tau+\ell)$, $u^{\text{ref}}_{\tau+\ell}:=u_{\text{ref}}(\tau+\ell)$, $\hat y^{P}_{\tau+\ell}:=\hat y^{P}(\tau+\ell)$, $ y_{\tau+\ell}:=y(\tau+\ell)$ $u_{\tau+\ell}:=u(\tau+\ell)$.

\subsubsection{Nonlinear case}\label{sec:NonlinearCase}

We begin by analyzing the NFIR case, where the cost \eqref{PosteriorAvgCost} can be easily derived in closed form. Then, we consider the NARX case, which requires a more careful analysis due to the dependence on past outputs of the predictors.\\

\noindent\textbf{NFIR.} The output data are generated as
\begin{equation}\label{NFIRMod}
        y(t) = f(u^t;\theta) + e(t),\\ 
\end{equation}
and the conditional predictors $\hat{y}^{P}_{\tau+\ell}=f(u^{\tau+\ell};\theta^{P})$, $\ell=0,\dots,T-1$, depend only on past inputs and $\theta^{P}$.
 
Since $\hat{y}^{P}_{t+\ell}$ does not depend on $E$, the cost \eqref{PosteriorAvgCost} can~be~written~as
\begin{align}
\hat{J}_u &= \sum_{\ell=0}^{T-1}\!   \int\! \|y^{\text{ref}}_{\tau+\ell}-\hat{y}^{P}_{\tau+\ell}\|^{2}_{Q_\ell}p(\theta^P)p(E) d\theta^P dE\!+\!    \|u^{\text{ref}}_{\tau+\ell}-u_{\tau+\ell}\|^{2}_{R_\ell}\notag\\
&= \sum_{\ell=0}^{T-1}\!   \int\! \|y^{\text{ref}}_{\tau+\ell}-\hat{y}^{P}_{\tau+\ell}\|^{2}_{Q_\ell}p(\theta^P) d\theta^P\!+\!    \|u^{\text{ref}}_{\tau+\ell}-u_{\tau+\ell}\|^{2}_{R_\ell}\notag\\
&= \sum_{\ell=0}^{T-1}\!   \| y^{\text{ref}}_{\tau+\ell}-\mathbb{E}[\hat{y}^{P}_{\tau+\ell}]\|^{2}_{Q_\ell} \!+\! \sum_{\ell=0}^{T-1} \tr Q_\ell\ \! \mathrm{Var}[\hat{y}^{P}_{t+\ell}] + \|u^{\text{ref}}_{\tau+\ell}-u_{\tau+\ell}\|^{2}_{R_\ell}\label{eq:costNFIR1}
\end{align} 
where in the last step the expectation and variance are taken w.r.t.~$\theta^P$ and we used the bias-variance decomposition
\begin{align*}
&\int \|y^{\text{ref}}_{\tau+\ell}-\hat{y}^{P}_{\tau+\ell}\|^{2}_{Q_\ell}p(\theta^P) d\theta^P \\
&= \int \|y^{\text{ref}}_{\tau+\ell}-\mathbb{E}[\hat{y}^{P}_{\tau+\ell}]-(\hat{y}^{P}_{\tau+\ell}-\mathbb{E}[\hat{y}^{P}_{\tau+\ell}])\|^{2}_{Q_\ell}p(\theta^P) d\theta^P\\
&= \|y^{\text{ref}}_{\tau+\ell}-\mathbb{E}[\hat{y}^{P}_{\tau+\ell}]\|^{2}_{Q_\ell}+\!\underbrace{\int \|\hat{y}^{P}_{\tau+\ell}-\mathbb{E}[\hat{y}^{P}_{\tau+\ell}]\|^{2}_{Q_\ell}p(\theta^P) d\theta^P}_{=\tr Q_\ell \mathrm{Var}[\hat{y}^{P}_{t+\ell}]}\\
&-2(y^{\text{ref}}_{\tau+\ell}-\mathbb{E}[\hat{y}^{P}_{\tau+\ell}])^\top Q_\ell \underbrace{\int(\hat{y}^{P}_{\tau+\ell}-\mathbb{E}[\hat{y}^{P}_{\tau+\ell}])p(\theta^P) d\theta^P}_{=0}.
\end{align*}
In the scalar Gaussian regression framework, we have $\tr Q_\ell = Q_\ell$ and $\mathbb{E}[\hat{y}^{P}_{\tau+\ell}]$, $\mathrm{Var}[\hat{y}^{P}_{t+\ell}]$ are given, respectively, by $\mathbb{E}[f(z_\ell^*)\,|\,\mathcal{D}]$, $\mathrm{Var}[f(z_\ell^*)\,|\,\mathcal{D}]$ in Proposition \ref{Prop1} with $z_{\ell}^{*}:=[u^{\tau+\ell}]^\top$. Thus, the cost \eqref{eq:costNFIR1} admits the closed-form expression
\begin{align}
\hat{J}_u &= \sum_{\ell=0}^{T-1}  Q_\ell\, \| y^{\text{ref}}_{\tau+\ell}-\mathbb{E}[f(z_\ell^*)\,|\,\mathcal{D}]\|^{2} + \sum_{\ell=0}^{T-1}R_\ell\,   \|u^{\text{ref}}_{\tau+\ell}\!-\!u_{\tau+\ell}\|^{2}\notag\\ &+\sum_{\ell=0}^{T-1}\!  Q_\ell   \mathrm{Var}[f(z_\ell^*)\,|\,\mathcal{D}]\,. \label{eq:costYNFIR2}
\end{align} 
Note that the previous expression is the sum of two terms: $$J_1:=\sum_{\ell=0}^{T-1}   \| y^{\text{ref}}_{\tau+\ell}-\mathbb{E}[f(z_\ell^*)\,|\,\mathcal{D}]\|^{2}_{Q_\ell}+\sum_{\ell=0}^{T-1}\!  \|u^{\text{ref}}_{\tau+\ell}\!-\!u_{\tau+\ell}\|^{2}_{R_\ell}$$ 
which can be interpreted as a nominal or ``certainty equivalence'' cost, and  $$J_2:=\sum_{\ell=0}^{T-1}\!  Q_\ell \mathrm{Var}[f(z_\ell^*)\,|\,\mathcal{D}]$$ 
which acts as a regularizer taking into account the effect on the cost of the uncertainty in predicting future outputs.\\

\noindent\textbf{NARX.}  The measurement model is as in \eqref{StochMod}, where we recall that $y^{t-1}$ and $u^{t}$ are understood as truncated to their first $m$ components.  

In this case, the conditional predictors $\hat{y}^{P}_{\tau+\ell}$ depend also on past outputs, and therefore on $E$. It is convenient to replace the dependence of the cost \eqref{PosteriorAvgCost} on $\theta^P$ and $E$ by a dependence on $\theta^P$ and the future conditional outputs $y^P_{\tau},y^P_{\tau+1},\dots$ contained in $Y^{P}$.  This is possible since $(\theta^{P},E)$ and  $(\theta^{P},Y^P)$ generate the same $\sigma$-algebra. As a consequence, letting $Y^P_{\ell}:=[y^P_{\tau},\dots,y^P_{\tau+\ell-1}]^\top$, the cost \eqref{PosteriorAvgCost} can be written as 
\begin{align}
\hat{J}_u &= \sum_{\ell=0}^{T-1}\!   \int\! \|y^{\text{ref}}_{\tau+\ell}-\hat{y}^{P}_{\tau+\ell}\|^{2}_{Q_\ell}p(\theta^P,Y^{P}_{\ell}) d\theta^P dY^{P}_{\ell}\!+\!    \|u^{\text{ref}}_{\tau+\ell}-u_{\tau+\ell}\|^{2}_{R_\ell}\notag\\
&= \sum_{\ell=0}^{T-1}\!   \int\! \|y^{\text{ref}}_{\tau+\ell}-\hat{y}^{P}_{\tau+\ell}\|^{2}_{Q_\ell}p(\theta^P|Y^{P}_{\ell})p(Y^{P}_{\ell}) d\theta^P dY^{P}_{\ell}\!+\! \|u^{\text{ref}}_{\tau+\ell}-u_{\tau+\ell}\|^{2}_{R_\ell}\notag\\
&= \sum_{\ell=0}^{T-1}\!   \int\! \| y^{\text{ref}}_{\tau+\ell}-\mathbb{E}[\hat{y}^{P}_{\tau+\ell}|Y^{P}_{\ell}]\|^{2}_{Q_\ell} p(Y^{P}_{\ell}) dY^{P}_{\ell}+\|u^{\text{ref}}_{\tau+\ell}-u_{\tau+\ell}\|^{2}_{R_\ell}\notag\\ 
&+ \sum_{\ell=0}^{T-1} \tr Q_\ell\!\! \int\!\! \mathrm{Var}[\hat{y}^{P}_{t+\ell}|Y^{P}_{\ell}] p(Y^{P}_{\ell}) dY^{P}_{\ell} \label{eq:costY}
\end{align} 
where in the last step the expectation and variance are taken w.r.t.~$\theta^P$ conditional on $Y^{P}_{\ell}$ and we performed a bias-variance decomposition of the term $\int\! \|y^{\text{ref}}_{\tau+\ell}-\hat{y}^{P}_{\tau+\ell}\|^{2}_{Q_\ell}p(\theta^P|Y^{P}_{\ell})d\theta^P$, similarly as before.

In the scalar Gaussian regression framework, \eqref{eq:costY} becomes
\begin{align}
\hat{J}_u &= \sum_{\ell=0}^{T-1}   \int \| y^{\text{ref}}_{\tau+\ell}-\mathbb{E}[f(z_\ell^*)\,|\,\mathcal{D},Y^{P}_{\ell}]\|^{2}_{Q_\ell} p(Y^{P}_{\ell}) dY^{P}_{\ell}\notag\\ 
&+ \sum_{\ell=0}^{T-1}\!  Q_\ell\!\! \int\!\! \mathrm{Var}[f(z_\ell^*)|\mathcal{D},Y^{P}_{\ell}] p(Y^{P}_{\ell}) dY^{P}_{\ell} +\sum_{\ell=0}^{T-1}\!  \|u^{\text{ref}}_{\tau+\ell}\!-\!u_{\tau+\ell}\|^{2}_{R_\ell}\!\label{eq:costYgaussian}
\end{align} 
where the terms $\mathbb{E}[f(z_\ell^*)\,|\,\mathcal{D},Y^{P}_{\ell}]$, $\mathrm{Var}[f(z_\ell^*)\,|\,\mathcal{D},Y^{P}_{\ell}]$, with $z_{\ell}^{*}:=[Y^P_\ell \  y^{\tau-1} \ u^{\tau+\ell}]^\top$, can be computed by updating the expressions $\mathbb{E}[f(z_\ell^*)\,|\,\mathcal{D}]$ and $\mathrm{Var}[f(z_\ell^*)\,|\,\mathcal{D}]$ in Proposition \ref{Prop1} through Kalman filter-like formulas. In fact, to account also for the additional conditioning on $Y^{P}$, one does not have to perform calculations from scratch. The current posterior covariance can be extended over the new measured input locations and the posterior of $f$ can then be updated with complexity cubic in the control horizon $T$, see, e.g., \cite{PillonettoDCN:08}. In practice, however, the size of $\mathcal{D}$ is typically much larger than the control horizon $T$, and $\mathbb{E}[f(z_\ell^*)\,|\,\mathcal{D},Y^{P}_{\ell}]$, $\mathrm{Var}[f(z_\ell^*)\,|\,\mathcal{D},Y^{P}_{\ell}]$ can be approximated by $\mathbb{E}[f(z_\ell^*)\,|\,\mathcal{D}]$, $\mathrm{Var}[f(z_\ell^*)\,|\,\mathcal{D}]$, respectively, allowing for the direct use of the expressions in Proposition \ref{Prop1}.

Beyond this harmless approximation, we introduce the following approximate input locations to obtain a closed-form expression for the cost
\begin{align*}
\tilde{z}_0^*&= z_0^*\\
\tilde{z}_1^*&= [\mathbb{E}[f(\tilde{z}_0^*)\,|\,\mathcal{D}] \  y^{\tau-1} \ u^{\tau+1}]^\top\\
\tilde{z}_2^*&= [\mathbb{E}[f(\tilde{z}_1^*)\,|\,\mathcal{D}]\  \mathbb{E}[f(\tilde{z}_0^*)\,|\,\mathcal{D}] \  y^{\tau-1} \  u^{\tau+2}]^\top\\
& \ \, \vdots\\
\tilde{z}_{T-1}^*&= [\mathbb{E}[f(\tilde{z}_{T-2}^*)\,|\,\mathcal{D}]\ \ldots \mathbb{E}[f(\tilde{z}_0^*)\,|\,\mathcal{D}] \  y^{\tau-1} \ u^{\tau+T-1}]^\top,
\end{align*}
where we stress that the vector
$$[\mathbb{E}[f(\tilde{z}_{\ell}^*)\,|\,\mathcal{D}]\  \ldots \ \mathbb{E}[f(\tilde{z}_0^*)\,|\,\mathcal{D}] \  y^{\tau-1}]^{\top},\ \ \  \ell=0,\dots,T-1,$$ should be interpreted as truncated at the first $m$ components when used as argument of $f(\cdot)$.

Using the approximate input locations above, the integrals in \eqref{eq:costYgaussian}  w.r.t.~$Y_\ell^P$ vanish, and the resulting (approximate) MPC cost is given by
\begin{align}
\hat{J}_u &= \sum_{\ell=0}^{T-1}  Q_\ell\, \| y^{\text{ref}}_{\tau+\ell}-\mathbb{E}[f(\tilde z_\ell^*)\,|\,\mathcal{D}]\|^{2} +\sum_{\ell=0}^{T-1}\!  R_\ell\,\|u^{\text{ref}}_{\tau+\ell}\!-\!u_{\tau+\ell}\|^{2}\notag \\
&+ \sum_{\ell=0}^{T-1}  Q_\ell \mathrm{Var}[f(\tilde z_\ell^*)|\mathcal{D}] \,.\label{eq:costYgaussian_approx}
\end{align} 
Similarly to the NFIR case, the previous expression can be seen as the sum of a nominal or ``certainty equivalence'' term
\begin{equation}\label{J1}
J_1 :=\sum_{\ell=0}^{T-1}  Q_\ell\, \| y^{\text{ref}}_{\tau+\ell}-\mathbb{E}[f(\tilde z_\ell^*)\,|\,\mathcal{D}]\|^{2} +\sum_{\ell=0}^{T-1}\!  R_\ell\,\|u^{\text{ref}}_{\tau+\ell}\!-\!u_{\tau+\ell}\|^{2}
\end{equation}
and a regularization term 
\begin{equation}\label{J2}
J_2:=\sum_{\ell=0}^{T-1}  Q_\ell \mathrm{Var}[f(\tilde z_\ell^*)|\mathcal{D}]
\end{equation}
accounting for the uncertainty around the predictors.

\subsubsection{Linear case}

Consider now systems that are linear in the past inputs and outputs as in \eqref{StochModLin}. For such systems, explicit expressions for both the cost and the optimal input sequence can be derived. 

Let $Y$ denote the vector with the future outputs $y_{\tau},\dots,y_{\tau+T-1}$ which we would like to drive to
$y^{\text{ref}}_{\tau},\dots,y^{\text{ref}}_{\tau+T-1}$. For our purposes,
using the formulas in \cite[Sec.~3.2]{Ljung:99}, it is important to write
the measurements model as follows
\begin{equation}\label{ModMeasMultiple}
Y= A(\theta) U + B(\theta) Y_{-} +  C(\theta) U_{-} + D(\theta)E
\end{equation}
where the entries of matrices $A$, $B$, $C$, $D$ are nonlinear functions of $\theta$ such that the $k$th row of $A(\theta) U + B(\theta) Y_{-} +  C(\theta) U_{-}$
is the $k$-step ahead predictor of $y_{\tau+k-1}$. In addition, the vectors $U_{-}$, $Y_{-}$ contain the past inputs $u_{\tau-1},u_{\tau-2},\dots$, and outputs $y_{\tau-1},y_{\tau-2},\dots$ while $U\in\mathbb{R}^{T}$ is the vector containing the inputs to be optimized, namely  $u_{\tau},\dots,u_{\tau+T-1}$. Conditional on the identification data $\mathcal{D}$, the model becomes
$$
Y^P = A(\theta^P) U + B(\theta^P) Y_{-} +  C(\theta^P) U_{-} + D(\theta^P)E.
$$
It holds
\begin{align}\label{eq:lin-eq-1}
    & \mathbb{E}[\| Y_{\text{ref}} - Y^P \|^2 ] = \notag \\
    & \mathbb{E}[\| Y_{\text{ref}} - A(\theta^P) U - B(\theta^P) Y_{-} -  C(\theta^P) U_{-} \|^2] + \mathbb{E}[\| D(\theta^P)E\|^2] \notag \\
    & + 2 \mathbb{E}\Big[ \sum_i [A(\theta^P) U + B(\theta^P) Y_{-} +  C(\theta^P) U_{-}]_i [D(\theta^P)E]_i \Big]
\end{align}
where $Y_{\text{ref}}\in\mathbb{R}^{T}$ contains the reference outputs  
$y^{\text{ref}}_{\tau},\dots,y^{\text{ref}}_{\tau+T-1}$. 
Note that the second term on the right-hand side of \eqref{eq:lin-eq-1} does not depend on $U$ 
and the last term is zero since the zero-mean noise $E$ is independent of $\theta^P$. This, combined with Remark \ref{RemIrrNoise}, allows us to write the conditional MPC cost as  
 \begin{equation}
  J_u^P  =   \|Y_{\text{ref}} -  A(\theta^P) U - B(\theta^P) Y_{-} -  C(\theta^P) U_{-}\|_{Q}^2 +  \|U_{\text{ref}} -  U\|_{R}^2 \label{eq:quadraticcost}
 \end{equation}
where $Q:=\mathrm{blkdiag}\{Q_\ell\}_{\ell=0,\dots,T-1}$, $R:=\mathrm{blkdiag}\{R_\ell\}_{\ell=0,\dots,T-1}$ while $U_{\text{ref}}$ contains the reference inputs $u^{\text{ref}}_{\tau},\dots,u^{\text{ref}}_{\tau+T-1}$. 
The posterior mean of \eqref{eq:quadraticcost} is
  \begin{align*}
  \hat{J}_u   & =   U^{\top}\mathbb{E}[A^{\top}Q A\,|\, \mathcal{D}]U -2 U^{\top}\!  (\mathbb{E}[A^{\top}|\, \mathcal{D}]QY_{\text{ref}} -  \mathbb{E}[A^{\top}\! QB\,|\, \mathcal{D}]Y_{-}\! \\ & -  \mathbb{E}[A^{\top}\!QC\,|\, \mathcal{D}] U_{-})  +  U^{\top} RU - 2  U^{\top}RU_{\text{ref}}  + H,
 \end{align*}
 where $H$ is a term not depending on $U$. From the above expression, it follows that the sequence of optimal inputs is
 \begin{align}\label{OptimalU} \nonumber
 U^{*} =&\  (\mathbb{E}[A^{\top}QA\,|\, \mathcal{D}]+R)^{-1}(\mathbb{E}[A^{\top}\,|\, \mathcal{D}]QY_{\text{ref}}\, - \\ 
 &\ \mathbb{E}[A^{\top} QB\,|\, \mathcal{D}]Y_{-} -  Q\mathbb{E}[A^{\top}QC\,|\, \mathcal{D}] U_{-}+R U_{\text{ref}}).
 \end{align}
Even if the entries of matrices $A$, $B$, $C$ are nonlinear functions of the random vector $\theta^P$, 
the terms $\mathbb{E}[A^{\top}Q A\,|\, \mathcal{D}]$, $\mathbb{E}[A^{\top}|\, \mathcal{D}]$, $\mathbb{E}[A^{\top}\!C\,|\, \mathcal{D}]$ appearing in $U^*$ can be computed offline either in closed-form or numerically via a simple Monte Carlo strategy. In particular, such Monte Carlo strategy involves generating a set of samples of $\theta^P$ based on its distribution described by \eqref{EandV1Lin} and then averaging the evaluated terms across the samples to approximate their expected values.

\section{Bayesian separation principle in action: linear setting}

The importance of the Bayesian separation principle
is now illustrated in the linear case, introducing
a simple yet insightful example.

\subsection{Closed-form MPC cost in a simple linear setting}

Consider a scalar output $y(t) \in \mathbb{R}$ and 
the reference trajectory $Y_{\text{ref}} \in \mathbb{R}^T$.
Following the notation of the previous section,
the predictor of $Y_{\text{ref}}$ is 
$$
A(\theta)U + B(\theta) Y_- +C(\theta) U_-
$$
where $A \in \mathbb{R}^{T \times T}$, $B \in \mathbb{R}^{T \times m}$,
$C \in \mathbb{R}^{T \times m-1}$
and the vector $U$ contains the $T$ optimization variables.
Consider the simplest case where the system memory is just $m=1$, which implies
$\theta \in \mathbb{R}^2$.
The one-step ahead predictor is 
\begin{equation}\label{FirstPred}
\hat{y}_1 = \theta_1 u_1 + \theta_2 y_0,
\end{equation}
the two-step ahead predictor is
\begin{equation}\label{SecondPred}
\hat{y}_2 = \theta_1 u_2 + \theta_2 \theta_1 u_1 + \theta_2^2 y_0
\end{equation}
and so on.
More in general, one has
$$
A=\left[\begin{array}{ccccc}\theta_1 & 0& 0 & 0 & \ldots \\
\theta_2\theta_1  & \theta_1 & 0 &  0 & \ldots  \\\theta_2^2\theta_1  & \theta_2\theta_1  & \theta_1 & 0 & \ldots 
\\ \ldots & \ldots & \ldots & \ldots & \ldots \end{array}\right],
$$
$$
B=\left[\begin{array}{c}\theta_2 \\\theta_2^2 \\\theta_2^3 \\ \vdots \end{array}\right]
$$
while $C$ is the null matrix.

To further simplify the example let $T=2$, hence only the predictors \eqref{FirstPred} and \eqref{SecondPred} will enter into the MPC cost. One has 
\begin{align*}
A^\top A &= \left[\begin{array}{cc}\theta_1^2+\theta_1^2\theta_2^2  & \theta_1^2 \theta_2  \\ \theta_1^2 \theta_2 & \theta_1^2\end{array}\right],\\
B^\top B &= \theta_2^2 + \theta_2^4, \\
A^\top B &= \left[\begin{array}{cc}\theta_1\theta_2+\theta_2^3\theta_1  \\ 
\theta_1 \theta_2^2 \end{array}\right].
\end{align*}
After seeing the identification data $\mathcal{D}$, we assume that 
the vector $\theta$ is Gaussian with mean $\mu$ with components $\mu_1, \mu_2$. Its covariance $\Sigma$ has $(i,j)$ entry denoted by $\sigma_{ij}$ outside the diagonals and by
$\sigma_i^2$ along the diagonal. 
After simple calculations based on non-central Gaussian moments,
one then obtains
\begin{align*}
\mathbb{E}[A^\top | \mathcal{D}] &= \left[\begin{array}{cc}
\mu_1  & \mu_1\mu_2 + \sigma_{12} \\ 
0   & \mu_1
\end{array}\right],
\end{align*}
where
\begin{align*}
\left[\mathbb{E}[A^\top A | \mathcal{D}] \right]_{11} &=
\mu_1^2(\mu_2^2 + 1) + \sigma_1^2 (\sigma_2^2+ 1)+\sigma_1^2\mu_2^2 \\
&\ \ \ \  + \sigma_2^2\mu_1^2 +2 \sigma_{12}^2
+4 \mu_1\mu_2 \sigma_{12},\\
\left[\mathbb{E}[A^\top A  | \mathcal{D}] \right]_{12} &= [\mathbb{E}[A^\top A]]_{21} =  \mu_1^2\mu_2+2\mu_1\sigma_{12} +\mu_2\sigma_1^2,\\
\left[\mathbb{E}[A^\top A | \mathcal{D}] \right]_{22} &= \mu_1^2 + \sigma_1^2, 
\end{align*}
and
\begin{align*}
&\mathbb{E}[A^\top B | \mathcal{D}] \\
&= \left[\begin{array}{cc}\mu_1\mu_2+\sigma_{12}+
3\sigma_2^2\sigma_{12}+3\mu_2^2\sigma_{12}+\mu_1(\mu_2^3+3\mu_2\sigma_2^2)\\ 
\mu_2^2\mu_1+2\mu_2\sigma_{12}+\sigma_2^2\mu_1 \end{array}\right].
\end{align*}

\subsection{Numerical experiment}

\begin{figure*}
\center {\includegraphics[scale=0.35]{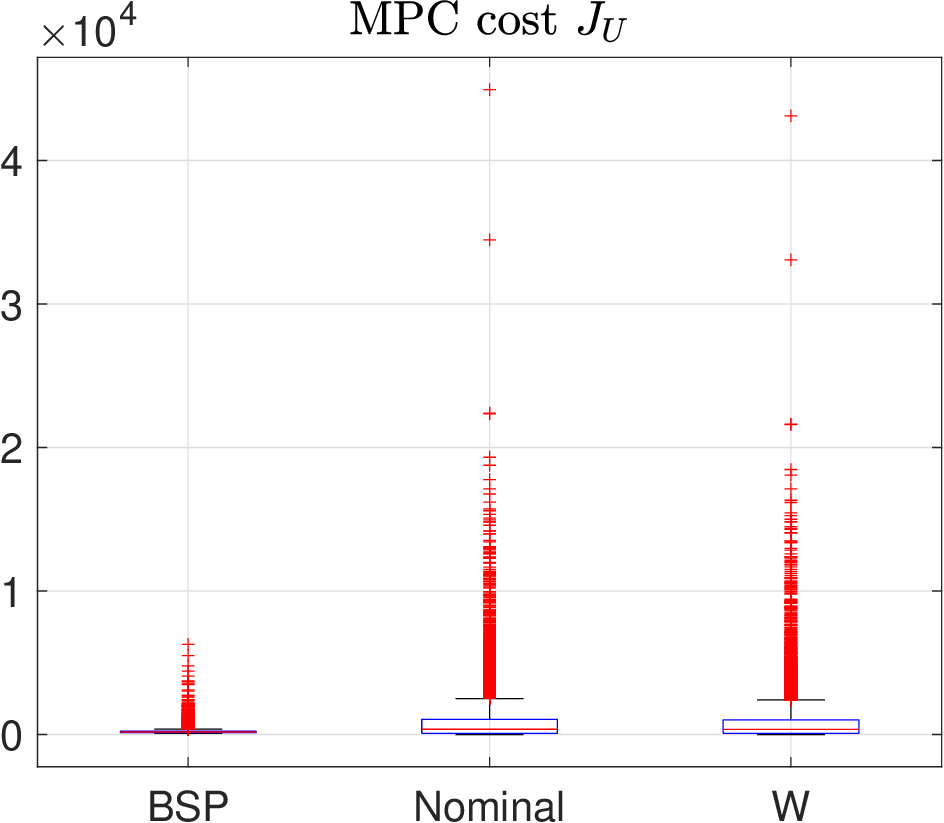}}  \ \ 
{\includegraphics[scale=0.35]{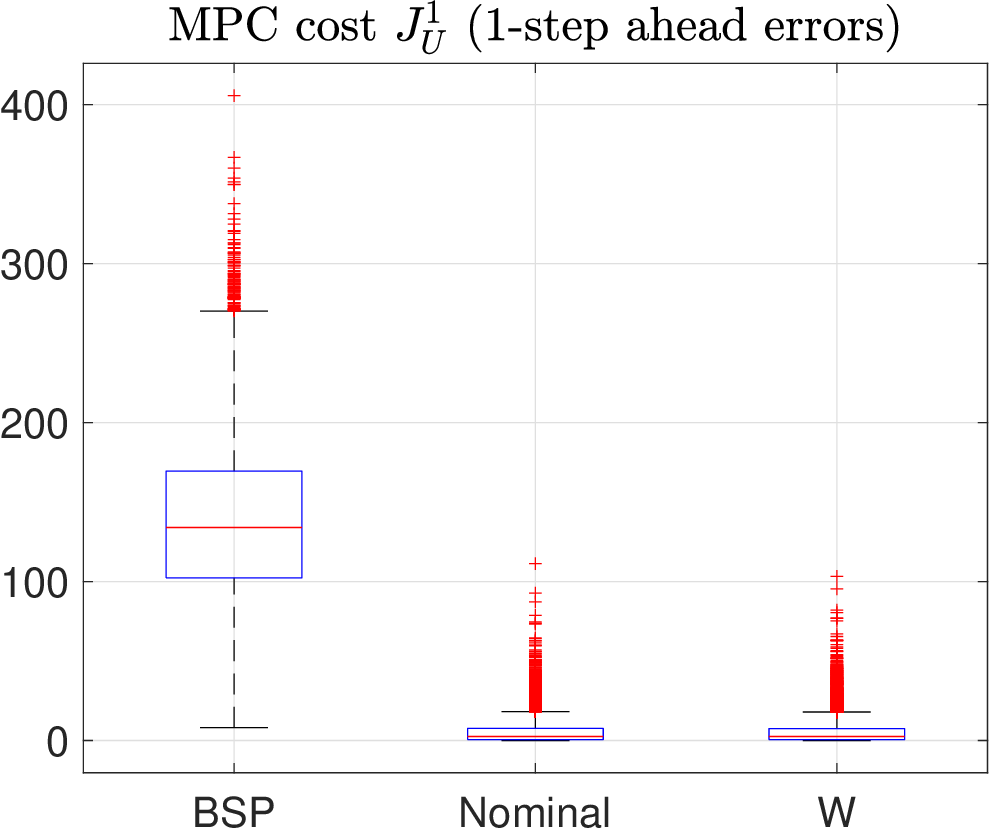}} \ \
{\includegraphics[scale=0.35]{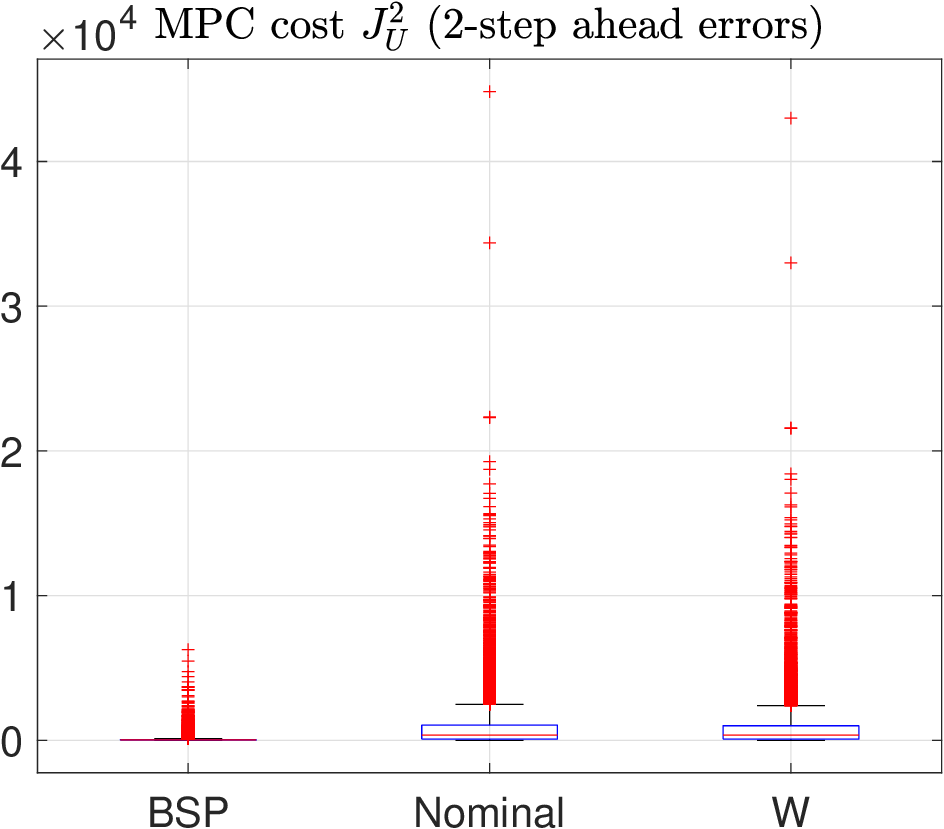}} 
\caption{{\bf{Left}}: boxplot of 10000 costs returned by MPC using only the
nominal model ({\bf{Nominal}}), exploiting the Bayesian separation principle to compute the input which minimizes the minimum variance estimate of the cost ({\bf{BSP}}) or introducing a special (stochastic) weight matrix 
({\bf{W}}). {\bf{Center}}: boxplot of the 10000 costs related to the part of the objective containing the one-step ahead prediction error. {\bf{Right}}: boxplot of the 10000 costs related to the part of the objective containing the two-step ahead prediction error.}
\label{Fig1}
\end{figure*}

Let
$$
\mu= \left[\begin{array}{cc}10  \\ 
5 \end{array}\right], \quad \Sigma= \left[\begin{array}{cc} 
4 & 0.9 \\ 
0.9 & 4 \end{array}\right], \quad Y_{\text{ref}}= \left[\begin{array}{cc} 10  \\ 
10 \end{array}\right]
$$
and $Y_-=1$.
The MPC cost \eqref{ControlCostBayes} is defined setting $Q$ to the identity matrix  and $R$ to the null matrix, hence no penalty on the control input is assigned. A Monte Carlo of 10000 runs is performed. At any run the system is obtained by drawing $\theta$ from the Gaussian distribution of mean $\mu$ and
covariance $\Sigma$. Let us consider three different control schemes:
\begin{itemize}
\item {\bf{Oracle:}} the true system, i.e. the realization of $\theta$, is known and used to build the optimal one- and two-step ahead predictors. This defines the optimal target as $J_u=0$.
In fact, $A$ is full-rank with probability one, hence
one can find $U$ that interpolates perfectly $Y_{\text{ref}}$; 
\item {\bf{Nominal:}} only the system estimate $\mu$ is used.
So, the uncertainty around the predictor is neglected:
the closed-form formulas obtained in this section are plugged 
in \eqref{OptimalU} with $\Sigma$ set to the null matrix.
The control input is the same at any run, given by
\begin{equation}\label{NomInput}
\hat{U}^N = \left[\begin{array}{c} 0.5 \\-4 \end{array}\right];
\end{equation}
\item {\bf{BSP:}} the Bayesian separation principle is used.
The closed-form formulas reported above are used in \eqref{OptimalU}, hence 
obtaining the input which minimizes the minimum variance estimate of the MPC cost. The control input is now given by
\begin{equation}\label{BSPInput}
\hat{U}^{BSP} \approx \left[\begin{array}{c} -0.65 \\ 1.48 \end{array}\right].
\end{equation}
\end{itemize}
At any run, the controls \eqref{NomInput} and \eqref{BSPInput}
are applied to the true system and the 
related MPC cost is evaluated. One thus obtains 10000 MPC costs from 
{\bf{Nominal}} and 10000 from {\bf{BSP}}.
For illustration purposes it is useful also to decompose the MPC objective into two parts $J^1_u$ and $J^2_u$ which describe the one- and two-step ahead error prediction:
$$
J_u(U) = J^1_u(U) + J^2_u(U)
$$
where
$$
J^i_u(U) = (Y_{\text{ref}}(i)-[AU]_i-[BY_-]_i)^2, \quad i=1,2.
$$
Fig. \ref{Fig1} (left panel) reports the boxplots of the MPC cost $J_u$.
The average cost from {\bf{BSP}} is around 217, almost four times smaller than that returned by {\bf{Nominal}} which is around 935 (results from a third control strategy denoted by ${\bf{W}}$ will be described later on). To better understand what is happening, boxplots of the two costs $J^1_u$ and $J^2_u$ are also reported. 
From the center panel, it is interesting to note that {\bf{Nominal}} 
outperforms {\bf{BSP}} in one-step-ahead prediction. 
But the right panel shows the effect of the Bayesian separation principle: it sacrificed one-step performance to improve accuracy over two steps  exploiting information on predictor parameters uncertainty. As a result, the overall error is reduced.

\subsection{Comparison with the approach in \cite{AC-MF-VB-SF:25}}

Working in the linear setting, 
Theorem 1 in \cite[Section 4]{AC-MF-VB-SF:25} reports a closed-form expression of the MPC cost obtained using a special (stochastic) weight matrix $W$ depending on $\theta$. 
For known $\theta$, such matrix is such that   
$\sigma^2W^{-1}W^{-\top}$ is the covariance of the prediction errors. From \eqref{ModMeasMultiple}, the following correspondence is then immediately obtained
$$
D(\theta)=W^{-1}(\theta),
$$
and the objective minimized in \cite{AC-MF-VB-SF:25}
corresponds to the expectation of the following variation of
\eqref{eq:quadraticcost}:
 \begin{align}
    \nonumber 
  J_u^W &= \Big\|D^{-1}(\theta)\Big(Y_{\text{ref}} -  A(\theta^P) U - B(\theta^P) Y_{-} -  C(\theta^P) U_{-}\Big)\Big\|_{Q}^2 \\ \nonumber
  &\ \ \ \ +  \|U_{\text{ref}} -  U\|_{R}^2\\ \nonumber
  &= \Big\|\Big(Y_{\text{ref}} -  A(\theta^P) U - B(\theta^P) Y_{-} -  C(\theta^P) U_{-}\Big)\Big\|_{W^{\top}(\theta)QW(\theta)}^2 \\ \label{eq:quadraticcostW}
  &\ \ \ \ +  \|U_{\text{ref}} -  U\|_{R}^2.
 \end{align}

In our simple example one has 
$$
D(\theta) = \left[\begin{array}{cc}1 & 0 \\\theta_2 & 1\end{array}\right], \quad D^{-1}(\theta) = \left[\begin{array}{cc}1 & 0 \\-\theta_2 & 1\end{array}\right], 
$$
and with $Q=I$ and $R=0$ the cost \eqref{eq:quadraticcostW} becomes
\begin{equation}\label{eq:quadraticcostW2}
J_u^W = \left\|\left[\begin{array}{cc}1 & 0 \\-\theta_2 & 1\end{array}\right]Y_{\text{ref}}
-U \theta_1 - \left[\begin{array}{c}\theta_2 \\0\end{array}\right]Y_-
\right\|^2
\end{equation}
which, apart from terms independent of $U$, is equal to
$$
U^\top \left[\begin{array}{cc}\theta_1^2 & 0 \\0 & \theta_1^2\end{array}\right]U+ 2 U^\top 
\left( \left[\begin{array}{cc}-\theta_1 & 0 \\ \theta_1\theta_2 & -\theta_1 \end{array}\right]Y_{\text{ref}}
+ \left[\begin{array}{c}\theta_1\theta_2 \\0\end{array}\right]Y_- \right).
$$
The expected cost $\mathbb{E}[J_u^W]$
is calculated using
$$
\mathbb{E}[\theta_1| \mathcal{D}] = 10, \quad \mathbb{E}[\theta_2^2| \mathcal{D}] = 104, \quad  \mathbb{E}[\theta_1\theta_2| \mathcal{D}] = 50.9,
$$
and its minimizer defines the following control input 
\begin{equation}\label{WInput}
\hat{U} = \frac{1}{104}\left[\begin{array}{c} 49.1 \\-409 \end{array}\right] \approx \left[\begin{array}{c} 0.47 \\-3.93 \end{array}\right].
\end{equation}
It is very close to the input \eqref{NomInput} obtained using the nominal model. The outcomes are indeed very similar as visible from the boxplots of the 10000 MPC costs, labelled 
{\bf W} in Fig. \ref{Fig1}, obtained using \eqref{WInput} at any Monte Carlo run. 
One can conclude that the price of using the result in
Theorem 1 in \cite[Section 4]{AC-MF-VB-SF:25} is to include a stochastic weight matrix that favours small prediction horizons. In fact, 
as already said, once fixed $\theta$ the matrix $W^\top W$ is (proportional to) the inverse of the covariance of the prediction errors whose diagonal elements increase with the prediction horizon. In our example, one indeed has
$$
\mathbb{E}[W^\top W | \mathcal{D}] \approx \left[\begin{array}{cc} 29.48 & -4.95 \\ -4.95 & 1\end{array}\right].
$$
One could certainly use the matrix $Q$ (here placed at identity) to try to counterbalance this effect but its choice is not so obvious due to the random nature of $W$. Using our control strategy \eqref{OptimalU} the effect is instead always transparent, e.g. $Q$ set to the identity 
always weights the prediction horizons equally.

\section{Bayesian separation principle in action: nonlinear setting}
In this example, we show how the use of the uncertainty regularizer cost $J_2$, as discussed at the end of Section
\ref{sec:NonlinearCase}, enhances the performance of the MPC controller with respect to the standard MPC, in which the cost consists only of the ``certainty equivalence" term $J_1$. We compare the performance of the two controllers in two different numerical experiments.

Consider the controlled Van der Pol oscillator described by the dynamic equations
\begin{align*}
    &\dot{x}_1 = 2x_2, \\
    &\dot{x}_2 = -0.8x_1+2x_2-10x_1^2x_2+u, \\
    & y = x_2,
\end{align*}
where $u$ is the control signal and $y$ is the measured output. The models used in the two experiments are trained with one dataset each, denoted as $\mathcal{D}_1$ for the first experiment and $\mathcal{D}_2$ for the second experiment. To generate the datasets, in both the experiments we discretize the dynamics using the Runge-Kutta four method, with discretization time $T_s = 0.05s$, and simulate $100$ trajectories over a time horizon of $T = 3s$ per trajectory. The control signal is a random signal with uniform distribution over $[-1,1]$ for the dataset $\mathcal{D}_1$ and $[0,1]$ for the dataset $\mathcal{D}_2$. The models are estimated using the Gaussian Process Regression method over the model class of NARX models with memory of $m=4$ past outputs and inputs. 
To compute the regularized and standard MPC controllers we use the regularized cost $\hat{J}_u = J_1 + J_2$ and $\hat{J}_u = J_1$, where $J_1$ and $J_2$ are defined in  \eqref{J1} and \eqref{J2}, respectively. We set $Q_\ell = 1$ and $R_\ell = 0$ for all the values of the index $\ell$, and we consider a receding horizon of $T = 4$ samples. We consider, as output reference, the constant signal $y_{\text{ref}}(t) = 0$ for all $t$, i.e., we want to steer the system to the origin. 

The MPC problem for NARX models discussed in subsection \ref{sec:NonlinearCase} can be recast as a nonlinear MPC problem, given the knowledge of the predictor $\mathbb{E}[f(\cdot)|\mathcal{D}]$ and the variance $\mathrm{Var}[f(\cdot)|\mathcal{D}]$, as described next. Let 
$
x_\tau=\left[y^{\tau-1}\right]^T$,
$$
x_{\tau+\ell}=\left[
\begin{array}{c}
     \mathbb{E}[f(\tilde{z}_{\ell-1}^*)|\mathcal{D}]\\
     \vdots \\
     \mathbb{E}[f(\tilde{z}_0^*)|\mathcal{D}] \\
     y_{\tau-1}\\
     \vdots \\
     y_{\tau-m+\ell}
\end{array}
\right],\,\,\, \ell=1,\ldots, m-1,
$$
and
$$
x_{\tau+m}=\left[
\begin{array}{c}
     \mathbb{E}[f(\tilde{z}_{m-1}^*)|\mathcal{D}]\\
     \vdots \\
     \mathbb{E}[f(\tilde{z}_0^*)|\mathcal{D}]
\end{array}
\right].
$$
Then the dynamics for the state $x$ can be written as
\begin{align}\label{eq:state_space}
x_{\tau+\ell+1}&=\left[
\begin{array}{ccccc}
0 & 0 & &  &\\
1 & \ddots & \ddots& &\\
  & \ddots & &  &\\
  &&&& 0  \\
  &&& 1& 0
\end{array}
\right] \, x_{\tau+\ell} \, + 
\, \left[
\begin{array}{c}
\mathbb{E}[f(\tilde{z}_{\ell}^*)|\mathcal{D}]\\
  \vdots\\
  \\
  \vdots  \\
  0
\end{array}
\right]
\end{align}

The problem we aim to solve is
\begin{align*}
&\underset{u_\tau, \ldots, u_{\tau+T-1}}{\mathrm{argmin}} \,\,\,\hat{J}_u \\
&\text{subject to} \,\,\, \eqref{eq:state_space}
\end{align*}
where $\hat{J}_u$ is as in \eqref{eq:costYgaussian_approx}. This problem can be solved by several numerical toolboxes. 

\begin{figure}[h!]
    \center {\includegraphics[scale=0.45]{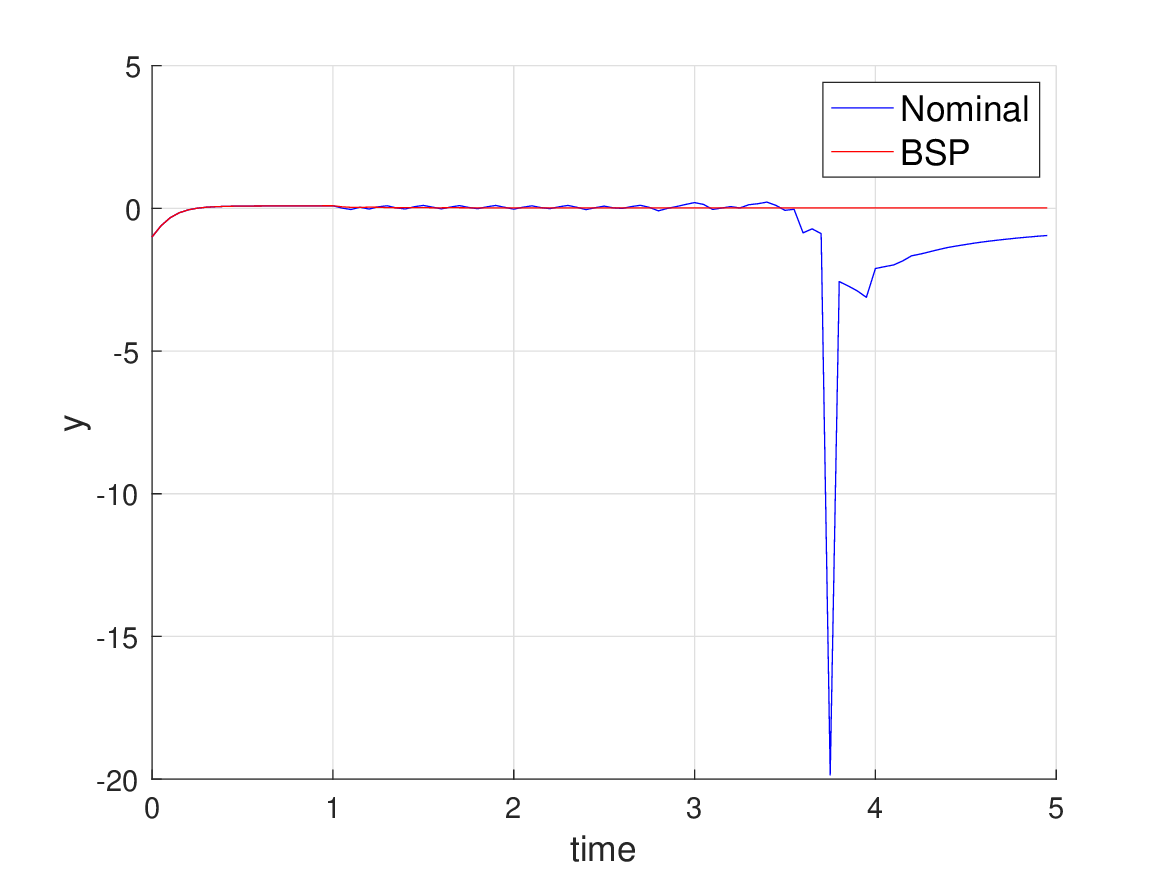}}  \ \ 
{\includegraphics[scale=0.45]{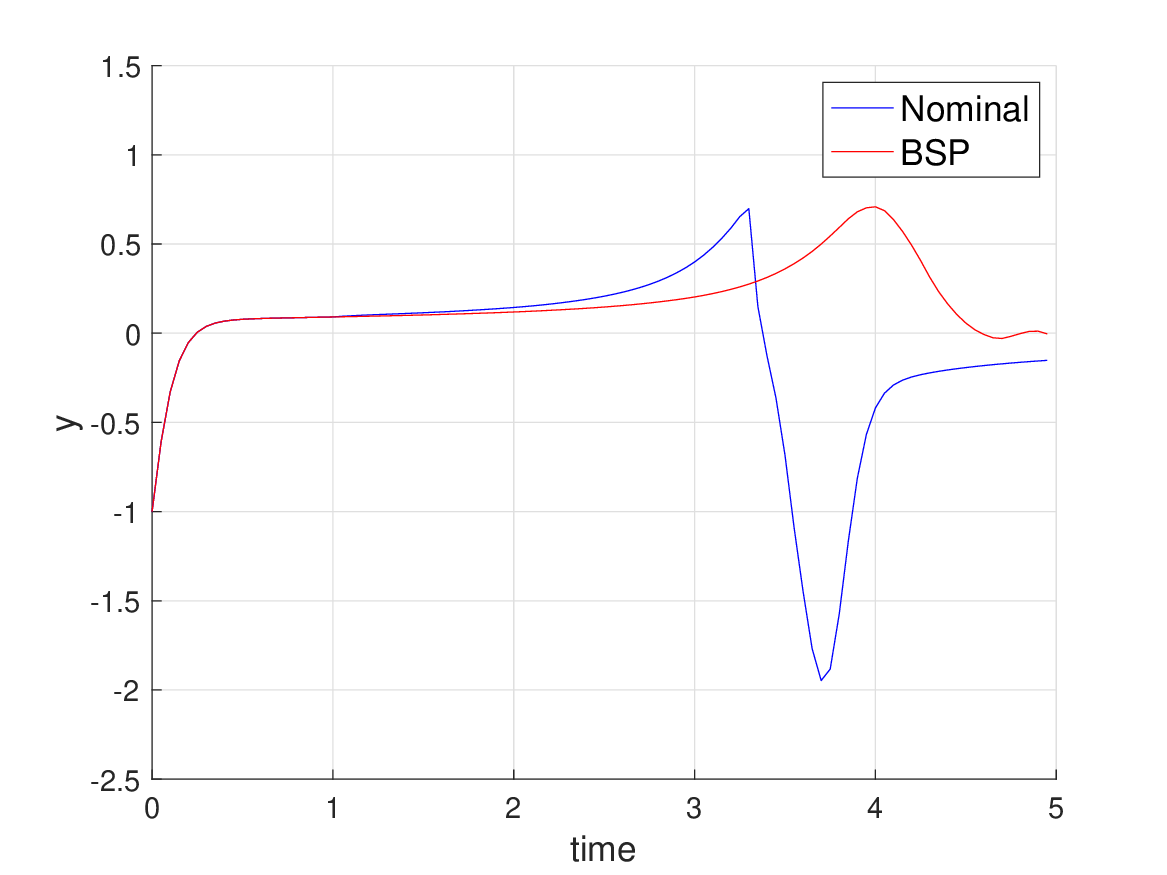}}
    \caption{Plots of the controlled $x_2$ trajectory using an MPC controller computed using the estimated model. {\bf{Top}}: Results of the first experiment, with the model trained using controls in the interval $[-1,1]$. The plot using the standard MPC is in blue, the plot using the MPC with regularized cost is in red. {\bf{Bottom}}: Results of the second experiment, with the model trained using controls in the interval $[0,1]$. The plot using the standard MPC is in blue, the plot using the MPC with regularized cost is in red.}
    \label{fig:numerical_example_J2}
\end{figure}

The results of the two experiments are shown in Figure~{\ref{fig:numerical_example_J2}}. 
Note that the results show important performance improvements when using the regularizer term $J_2$. 
    Indeed, in the first experiment, the model is trained over control signals in the interval $[-1,1]$. Approaching the time instant $3.5s$, the unforced system would show strong oscillations, and the output $y$ would go far away from the origin. To compensate for this behavior, the non-regularized MPC estimates an optimal control signal with a large magnitude, outside $[-1,1]$. However, these estimates are very unreliable, as the system is not trained with inputs in this interval, causing the poor performance shown in Figure~{\ref{fig:numerical_example_J2}}, top. The regularized MPC instead, thanks to the regularizer term $J_2$, prefers controls in $[-1,1]$, where the variance of the prediction is small. Although these may not be the optimal choices of the controls in the ideal case where the true system is known, they guarantee that the estimated model gives good predictions and, therefore, that the performance of the system does not degenerate.
    
    In addition, in the second experiment, a similar behavior is shown. The model is trained with inputs in the interval $[0,1]$. The non-regularized MPC, as before, estimates optimal controls outside $[0,1]$, causing the poor performance shown in Figure~{\ref{fig:numerical_example_J2}}, bottom. The regularized MPC, thanks to the regularizer term $J_2$, keeps the controls in the interval $[0,1]$. However, as discussed before, this is not in general the optimal choice for the true system. In particular, this restricts the choices of controls with respect to the first experiment, leading to the worse behavior shown to the right of Figure~{\ref{fig:numerical_example_J2}}, bottom.

\section{Conclusion}

In this paper, we explored the existence and role of a separation principle between identification and control design. By focusing on model predictive control, we argued that such separation principle holds asymptotically in the number of data in a Fisherian setting and also for finite data in a Bayesian setting. We further discussed how to cast MPC within a non-parametric Gaussian regression framework and how to exploit the separation principle to establish explicit, computationally tractable expressions for the control cost and optimal input sequence. The control inputs designed by leveraging this approach may exhibit superior performance compared to both nominal, uncertainty-agnostic techniques and alternative uncertainty-aware formulations, as illustrated by our numerical experiments. Throughout the paper we also provided precise definitions for commonly used terminology in the data-driven control literature.

Overall, this work thus points out important advantages related to the use of the Bayesian setting for control design. One could however argue that the price to pay is the need to introduce and believe in the prior. This aspect, however, is greatly mitigated by the introduction of very general priors typical of Gaussian regression which  include just regularity information on the input-output map. In addition, the hyperparameters that regulate the smoothness level are tuned from the data, providing a form of robustness. There are also routes to further enhance such robustness, e.g. resorting to the even more sophysticated frequentist Bayesian framework \cite{Berger,Baggio2022} which may account for misspecified priors when calculating uncertainty bounds. Possible developments of this work may thus include a comparison between different statistical paradigms for the design of controllers including Bayes, Fisher and frequentist Bayes.

\section*{Appendix: Proof of Proposition \ref{Prop1}}


The past input-output data preceding in time those contained in $\mathcal{D}$ 
are contained in the vector denoted by $z^-$ and are used to define the initial conditions of our dynamic models. We assume that 
the model \eqref{GRF} for $f$ is not influenced by these past input-output data. Therefore, in the following conditioning
on $z^-$ is omitted and the Gaussian prior on $f$ describes all system randomness before seeing only the data in $\mathcal{D}$. 
Since $u$ is a fixed deterministic input, we can also leave its dependence implicit and $Y$ can be used instead of $\mathcal{D}$ to define the $\sigma$-algebra on which $f$ is projected.

It is useful to define 
$$
\bar{f} = [f(z^1) \ \ldots \ f(z^N)]^\top,
$$
whose covariance coincides with the kernel matrix $\mathbb{K}$,
and the enriched  vector
$$
\bar{f}^* = [f(z^1) \ \ldots \ f(z^N) \ f(z^*)]^\top
$$
whose covariance is 
$$
C= \lambda \left[\begin{array}{cc}\mathbb{K} & \Gamma \\ \Gamma^\top & K(z^*,z^*)\end{array}\right].
$$
We also use $f^c$ to indicate the Gaussian random field over all the input locations except those
related to $\bar{f}^*$. This permits to interpret $f$ as the union of $\bar{f}^*$ and $f^c$.

The posterior 
$p(f|Y)$ is proportional to 
\begin{eqnarray*}
p(Y,f) &=& \left[ \prod_{i=1}^{N}
p(y_i|z^{i},f)\right] p(f)\\
 &=& \left[\prod_{i=1}^{N} p(y_i|f(z^{i})) \right] p(\bar{f}^*) p(f^c|\bar{f}^*)
\end{eqnarray*}
where we have used the chain rule. It is now immediate 
to integrate out $f^c$ from the above expression. In fact, the joint density 
of $Y$ and $\bar{f}^*$ is nothing more than 
$\left[\prod_{i=1}^{N} p(y_i|f(z^{i})) \right] p(\bar{f}^*)$.
One has that $Y -\bar{f}$ is white Gaussian noise
of variance $\sigma^2$, independent of $f$. Then, if $G$ indicates the $N \times (N+1)$ matrix with an identity matrix and a null column vector side by side, so that $\bar{f}=G\bar{f}^*$, it holds that
\begin{eqnarray}\label{MarginalPost}
-\log(p(\bar{f}^*,Y)) &=&  \frac{\| Y - 
G\bar{f}^*\|^2}{2 \sigma^2} + \frac{(\bar{f}^*)^\top C^{-1}\bar{f}^*}{2}\\
\nonumber &+&
\frac{1}{2}\log\det(4\pi^2 \sigma^2C).
\end{eqnarray}
Thus, $\bar{f}^*$ conditional on $Y$ remains Gaussian\footnote{It is also immediate to see that the same would be true if, instead of $f(z^*)$, we had added to $\bar{f}$ any other finite-dimensional vector extracted from $f$. So, this also shows that the entire Gaussian random field conditional on $Y$ remains Gaussian.} and standard calculations now lead to \eqref{EandV1} and \eqref{EandV2}.

As for the the marginal probability density function of $Y$ reported in \eqref{ML1}, it can be easily obtained using the Laplace integral to integrate out $\bar{f}^*$ from $p(\bar{f}^*,Y)$ \cite{Raftery}.
But an even simpler argument is to note that the minus log of the joint density reported in
\eqref{MarginalPost} is the same as if we had started from the model measurements
$Y=M\bar{f}^*+E$ where $\bar{f}^*$ is the Gaussian random field $f$ sampled on deterministic input locations that do not depend on $Y$ itself. This interpretation leads to a problem equivalent to the original one, just for the purposes of marginalization w.r.t $\bar{f}^*$, where the vector $Y$ is zero-mean Gaussian\footnote{This is certainly not true starting from our original model where $M$ also depends on $Y$, but again we emphasise that this is irrelevant for our integration problem.} with covariance $\Sigma_y=\lambda K+\sigma^2 I$. This immediately gives \eqref{MarginalPost} as the result of the integral.

Finally, in the linear case one uses 
the measurements model \eqref{StochModLin}
and the linear kernel \eqref{LinKer} which induces the following prior
on the vector $\theta$ of the predictor impulse responses coefficients:
$$
\theta \sim \mathcal{N}(0,\lambda M).
$$
Then, \eqref{EandV1Lin} is obtained just repeating all the arguments 
here exposed.


\begin{thebibliography}{34}
\setcounter{enumiv}{0}

\bibitem{Anderson:1979}
B.D.O. Anderson and J.B. Moore.
\newblock {\em Optimal Filtering}.
\newblock Prentice-Hall, Englewood Cliffs, N.J., USA, 1979.



\bibitem{Baggio2022}
G.~Baggio, A.~{Car{\`e}}, A.~Scampicchio  and G.~Pillonetto.
\newblock Bayesian frequentist bounds for machine learning and system
  identification.
\newblock {\em Automatica}, 146:110599, 2022.

\bibitem{Bell2004}
B.~Bell and G.~Pillonetto.
\newblock Estimating parameters and stochastic functions of one variable using nonlinear measurement models.
\newblock {\em Inverse Problems}, 20(627), 2004.

\bibitem{berberich2020data}
J.~Berberich, J.~K{\"o}hler, M.A.~M{\"u}ller and F.~Allg{\"o}wer
\newblock Data-driven model predictive control with stability and robustness guarantees.
\newblock {\em IEEE Transactions on Automatic Control}, 66(4):1702--1717, 2020.

\bibitem{Berger}
M. Bayarri and J. Berger
\newblock The Interplay of Bayesian and Frequentist Analysis.
\newblock {\em Stat Sci}, 19:58-80, 2004.





  \bibitem{Care}
A. Carè, R. Carli, A. Dalla Libera, D. Romeres and G. Pillonetto
\newblock Kernel methods and Gaussian processes for system identification and control: A road map on regularized kernel-based learning for control.
\newblock {\em IEEE Control Systems Magazine}, 43(5), 69-110, 2023.


\bibitem{CaseBerg:01} G.~Casella and R.~Berger \newblock {\em Statistical Inference}. 
\newblock Pacific Grove, CA, Thomson Learning, 2001.

\bibitem{Celi23}
F.~Celi, G.~Baggio and F.~Pasqualetti. 
\newblock Closed-form and robust expressions for data-driven LQ control.
\newblock {\em Annual Reviews in Control}, 56:100916, 2023.


\bibitem{AC-MF-VB-SF:25} 
A.~Chiuso, M.~Fabris, V.~Breschi and S.~Formentin. 
\newblock Harnessing uncertainty for a separation principle in direct data-driven predictive control. \newblock {\em Automatica}, 173:112070, 2025.

\bibitem{coulson2019data}
J.~Coulson, J.~Lygeros and F.~D{\"o}rfler.
\newblock Data-enabled predictive control: In the shallows of the DeePC.
\newblock {\em 18th European Control Conference (ECC)}, 307--312, 2019.


\bibitem{DeN1997}
G. {De Nicolao}, G Sparacino and C Cobelli.
\newblock Nonparametric input estimation in physiological systems: problems, methods, and case studies.
\newblock {\em Automatica }, 33(5):851--870, 1997.

\bibitem{DePersisTesi19}
C.~De Persis and P.~Tesi
\newblock Formulas for data-driven control: Stabilization, optimality, and robustness,
\newblock {\em IEEE Transactions on Automatic Control}, 65(3):909--924, 2019.


\bibitem{Dorfler2023}
F.~D{\"o}rfler.
\newblock Data-Driven Control: Part One of Two: A Special Issue Sampling from a Vast and Dynamic Landscape.
\newblock {\em IEEE Control Systems Magazine}, 43(5), 24-27, 2023.

\bibitem{Dorfler2023b}
F.~D{\"o}rfler.
\newblock Data-driven control: Part two of two: Hot take: Why not go with models?
\newblock {\em IEEE Control Systems Magazine}, 43(6), 27-31, 2023.

\bibitem{Dorfler2022bridging}
F.~D{\"o}rfler, J.~Coulson and I.~Markovsky. 
\newblock Bridging direct and indirect data-driven control formulations via regularizations and relaxations
\newblock {\em IEEE Transactions on Automatic Control}, 68(2), 883-897, 2022.

\bibitem{Efron1973}
B.~Efron and C.~Morris.
\newblock Stein's estimation rule and its competitors--an empirical {B}ayes
  approach.
\newblock {\em Journal of the American Statistical Association},
  68(341):117--130, 1973.




\bibitem{Goodwin1992}
G.C. Goodwin, M.~Gevers and B.~Ninness.
\newblock Quantifying the error in estimated transfer functions with
  application to model order selection.
\newblock {\em IEEE Transactions on Automatic Control}, 37(7):913--928, 1992.

\bibitem{Hastie01}
T.~J. Hastie, R.~J. Tibshirani and J.~Friedman.
\newblock {\em The Elements of Statistical Learning. Data Mining, Inference and
Prediction}.
\newblock Springer, Canada, 2001.

\bibitem{Hewing19}
L.~Hewing, J.~Kabzan and M.~N.~Zeilinger. 
\newblock {Cautious model predictive control using gaussian process regression} 
\newblock {\em IEEE Transactions on Control Systems Technology}, 28.6 (2019): 2736-2743.

\bibitem{Hewing20}
L.~Hewing, K.P.~Wabersich, M.~Menner, and M.~N.~Zeilinger. 
\newblock {Learning-based model predictive control: Toward safe learning in control} 
\newblock {\em Annual Review of Control, Robotics, and Autonomous Systems}, 3.1 (2020): 269-296.


\bibitem{Krishnan2021direct}
V.~Krishnan and F.~Pasqualetti
\newblock On direct vs indirect data-driven predictive control.
\newblock {\em 2021 60th IEEE Conference on Decision and Control (CDC)}, 736-741, 2021.

\bibitem{Leeb2005}
H. Leeb and B.M. Potscher.
\newblock Model selection and inference: facts and fiction.
\newblock {\em Econometric Theory}, 21(1), 2005.

\bibitem{Ljung:99}
L.~Ljung.
\newblock {\em System Identification - Theory for the User}.
\newblock Prentice-Hall, Upper Saddle River, N.J., 2nd edition, 1999.

\bibitem{Ljung:2014}
L. Ljung, G.C. Goodwin and J.C. Agüero.
\newblock Stochastic Embedding revisited: A modern interpretation.
\newblock {\em Proceedings of the 53rd IEEE Conference on Decision and Control}, 2014

\bibitem{MacKayNN}
D.J.C. MacKay.
Bayesian Interpolation 
\newblock {\em Neural Computation}, 4(3), 1992.

\bibitem{Mattsson2024}
P. Mattsson, F. Bonassi, V. Breschi, and T.B. Sch\"on. 
\newblock On the equivalence of direct and indirect data-driven predictive control approaches. 
\newblock {\em IEEE Control Systems Letters}, 8 (2024): 796-801.

 \bibitem{Minh:09}
H.Q. Minh, P. Niyogi and Y. Yao. Mercer Theorem, Feature Maps, and Smoothing, in \emph{Learning Theory. COLT 2006. Lecture Notes in Computer Science}, vol. 4005, Springer, 2006.

\bibitem{Neal1995}
R.M. Neal, 
\newblock {\em Bayesian learning for neural networks},
\newblock Springer, 1995.

 \bibitem{DeepSurvey2025}
G. Pillonetto, A. Aravkin, D. Gedon, L. Ljung, A.H. Ribeiro and T.B. Schön. 
Deep networks for system identification: A survey
\newblock {\em Automatica}, 171, 2025.
 

\bibitem{PillonettoDCNL:14}
G.~Pillonetto, F.~Dinuzzo, T.~Chen, G.~{De Nicolao} and L.~Ljung.
\newblock Kernel methods in system identification, machine learning and
  function estimation: A survey.
\newblock {\em Automatica}, 50, March 2014.

\bibitem{PillonettoDCN:08}
G.~Pillonetto, F.~Dinuzzo and G.~{De Nicolao}. 
\newblock Bayesian online multitask learning of Gaussian processes. 
\newblock {\em IEEE Transactions on Pattern Analysis and Machine Intelligence}, 32(2): 193-205, 2008.

\bibitem{Pillonetto:11nonlin}
G.~{Pillonetto}, M.~H. {Quang} and A.~{Chiuso}.
\newblock A new kernel-based approach for nonlinear system identification.
\newblock {\em IEEE Transactions on Automatic Control}, 56(12):2825--2840,
  2011.

\bibitem{SpringerRegBook2022}
G.~Pillonetto, T.~Chen, A.~Chiuso, G.~De Nicolao and L.~Ljung.
\newblock {\em Regularized System Identification}.
\newblock Springer, 2022.


\bibitem{PillonettoPNAS}
G Pillonetto and L Ljung, Full {B}ayesian identification of linear dynamic systems
  using stable kernels.
\newblock {\em Proceedings of the National Academy of
  Sciences USA} \textbf{120} (2023).

\bibitem{Poggio75}
T.~Poggio.
\newblock On optimal nonlinear associative recall.
\newblock {\em Biological Cybernetics}, 19(4):201--209, 1975.


\bibitem{Raftery}
R.E. Kass and A.E. Raftery.
\newblock {B}ayes Factors.
\newblock {\em J. Amer. Statist. Assoc.}, 96:773-795, 1995.


\bibitem{Rasmussen}
C.E. Rasmussen and C.K.I. Williams.
\newblock {\em {G}aussian Processes for Machine Learning}.
\newblock The MIT Press, 2006.

\bibitem{Scampicchio}
A. Scampicchio, A. Chiuso, S. Formentin and G. Pillonetto.
\newblock Bayesian kernel-based linear control design.
\newblock {\em Proceedings of 58th IEEE conference on decision and control (CDC)}, 2019.

\bibitem{Scholkopf01b}
B.~Sch\"{o}lkopf and A.J. Smola.
\newblock {\em Learning with Kernels: Support Vector Machines, Regularization,
  Optimization, and Beyond}.
\newblock (Adaptive Computation and Machine Learning). MIT Press, 2001.

\bibitem{SchoukensLjung:2019}
J.~Schoukens and L.~Ljung.
\newblock Nonlinear system identification -- a user-oriented roadmap.
\newblock {\em IEEE Control Systems Magazine}, 39(6):28--99, December 2019.

\bibitem{Scovel10}
C.~Scovel, D.~Hush, I.~Steinwart and J.~Theiler.
\newblock Radial kernels and their reproducing kernel {H}ilbert spaces.
\newblock {\em Journal of Complexity}, 26(6):641 -- 660, 2010.


\bibitem{Soderstrom}
T.~S{\"o}derstr{\"o}m and P.~Stoica.
\newblock {\em System Identification}.
\newblock Prentice-Hall, 1989.

\bibitem{Wahba:90}
G.~Wahba.
\newblock {\em Spline models for observational data}.
\newblock SIAM, Philadelphia, 1990.


\bibitem{Yang2005}
Y.~Yang
\newblock Can the Strengths of AIC and BIC Be Shared? A Conflict between Model Indentification and Regression Estimation
\newblock {\em Biometrika}, 92(4): 937--950, 2005.


\bibitem{Zhang2023}
J. Zhang, Y.~Yang and J. Ding
\newblock Information criteria for model selection
\newblock {\em Wiley Interdisciplinary Reviews: Computational Statistics}, 15(5), 2023.

\bibitem{Zhena1966}
P.W. Zehna.
\newblock Invariance of Maximum Likelihood
\newblock {\em The Annals of Mathematical Statistics}, 37, 744, 1966.



\end{thebibliography}
\end{document}